\documentclass[aps,prb,twocolumn,superscriptaddress,floatfix,showpacs,groupedaddress]{revtex4-1}
\usepackage{graphicx}
\usepackage{amsfonts}
\usepackage{amssymb}
\usepackage{amsmath}
\usepackage{txfonts}
\usepackage{lipsum}
\usepackage{color}
\DeclareMathOperator{\Tr}{Tr}

\DeclareMathOperator{\sgn}{sgn}

\DeclareMathOperator{\Real}{Re}
\DeclareMathOperator{\Imag}{Im}

\begin{document}
\title{Dynamical control of electron-phonon interactions with high-frequency light}
\author{C. Dutreix}
\author{M. I. Katsnelson}
\affiliation{Radboud University, Institute for Molecules and Materials, Heyendaalseweg 135, 6525AJ Nijmegen, The Netherlands}

%\date{\today}

\begin{abstract}
This work addresses the one-dimensional problem of Bloch electrons when they are rapidly driven by a homogeneous time-periodic light and linearly coupled to vibrational modes. Starting from a generic time-periodic electron-phonon Hamiltonian, we derive a time-independent effective Hamiltonian that describes the stroboscopic dynamics up to the third order in the high-frequency limit. This yields nonequilibrium corrections to the electron-phonon coupling that are controllable dynamically via the driving strength. This shows in particular that local Holstein interactions in equilibrium are corrected by nonlocal Peierls interactions out of equilibrium, as well as by phonon-assisted hopping processes that make the dynamical Wannier-Stark localization of Bloch electrons impossible. Subsequently, we revisit the Holstein polaron problem out of equilibrium in terms of effective Green functions, and specify explicitly how the binding energy and effective mass of the polaron can be controlled dynamically. These tunable properties are reported within the weak- and strong-coupling regimes since both can be visited within the same material when varying the driving strength. 
This work provides some insight into controllable microscopic mechanisms that may be involved during the multicycle laser irradiations of organic molecular crystals in ultrafast pump-probe experiments, although it should also be suitable for realizations in shaken optical lattices of ultracold atoms.
\end{abstract}

\maketitle

%%%
\section{Introduction}

A polaron is a fermionic quasiparticle that was introduced by Landau in a 1933 seminal paper to describe the trapping of an electron by the ionic distorsion it induces in a crystal [\onlinecite{landau1933electron}]. The self-trapping of such an electron was subsequently studied in the case of weak electron-phonon coupling by Pekar and Fr\"ohlich [\onlinecite{pekar1946autolocalization},\,\onlinecite{frohlich1954electrons}]. They showed that, within a continuum dielectric medium, a single electron can drag a phonon cloud along a slow motion without being trapped, thus resulting in a large polaron that propagates freely with an effective mass. By opposition, the polaron size becomes small - of the order of the lattice constant - in the regime of a strong electron-phonon coupling compared to the electron bandwidth. This situation depicted by Holstein, Lang and Firsov refers to a quasi-trapped polaron that propagates with an exponentially heavier effective mass [\onlinecite{holstein1959studies},\,\onlinecite{lang1962title}]. Importantly, all these polaron features were finally unified within a path-integral-based variational approach that allowed Feynman to characterize the binding energy and effective mass of Fr\"ohlich's polaron for all coupling strengths [\onlinecite{feynman1948space},\,\onlinecite{feynman1955slow}].

From the experimental perspective these quasiparticles were first identified in uranium dioxide as small polarons [\onlinecite{nagels1963electrical}]. Later, localized lattice distortions were pointed out to affect the Curie temperature of the ferromagnetic transition in perovskites, and to be involved in the colossal magnetoresistance of manganites [\onlinecite{millis1995double,zhao1996giant,alexandrov1999carrier,sharma2002oxygen,edwards2002ferromagnetism,hartinger2006polaronic}]. Whereas the phonons turn out to be crucial in the context of symmetry breaking phase transitions with for example structural Peierls dimerization and conventional BCS superconductivity [\onlinecite{bardeen1957theory},\,\onlinecite{bardeen1957microscopic}], their coupling to the charge carriers would also play a significant role in high-temperature superconductors [\onlinecite{alexandrov1996coherent,bianconi1996determination,lanzara2001evidence,lee2006interplay,gweon2006strong,takahashi2008superconductivity,chen2008superconductivity,kresin2009colloquium}], although the underlying microscopic pairing mechanism has not been clearly identified yet. Polaron physics was also seriously discussed in connection to organic molecular crystals with possible applications as field-effects transistors [\onlinecite{sundar2004elastomeric,takeya2007very,kawai2012characteristics}]. It was first thought that local electron-phonon interactions of Holstein type were sufficient to explain understand the physics of organic semiconductors. Nevertheless, experiments achieved in aromatic hydrocarbon crystals showed that nonlocal electron-phonon interactions are also involved in transport properties [\onlinecite{roberts1980temperature}], resulting in many studies that aimed to highlight the interplay between local and nonlocal electron-phonon interactions in these organic materials [\onlinecite{munn1985theory, munn1985theory3, zhao1994munn,PhysRevLett.89.275503,zoli2005nonlocal,PhysRevLett.96.086601,PhysRevLett.103.266601,PhysRevB.82.035208,Ciuchi:2011dn,Li:2011nr,Li:2013eu}].

On the other hand, the last years witnessed a growing interest inside the condensed matter community for out-of-equilibrium physics [\onlinecite{aoki2014nonequilibrium}]. With the development of ultrafast pump-probe spectroscopy, it became possible to study excitation and relation processes, as well as steady regimes in many-body systems [\onlinecite{joura2008steady,tsuji2008correlated,tsuji2009nonequilibrium,wall2011quantum}], leading to phenomena such as ultrafast time-scale induced superconductivity [\onlinecite{fausti2011light}] and symmetry-protected topological transitions [\onlinecite{oka2009photovoltaic,lindner2011floquet,carpentier2015topological,dutreix2016laser}]. This is quite naturally then that the poralon problem was revisited in this nonequilibrium context. For example, the electron-phonon coupling offers a dominant relaxation channel to the photo-excited quasiparticles of Mott insulators [\onlinecite{PhysRevLett.112.117801}]. It was also reported that quenching the Holstein coupling reduces the Coulomb interaction and enhances the production of doublons in the Mott insulating phase [\onlinecite{PhysRevB.88.165108}]. In order to get some insight into the nonequilibrium dynamics of such many-body phases, the real-time dynamics of a single electron in Holstein model has recently been studied [\onlinecite{PhysRevB.91.104302},\,\onlinecite{PhysRevB.91.104301}]. This highlights for instance what the electron transient dynamics is, from the time at which a DC electric field is turned on until the electron reaches a steady state with constant velocity thanks to energy dissipation through optical phonons [\onlinecite{vidmar2011nonequilibrium}], as predicted by Thornber and Feynman in 1970 [\onlinecite{thornber1970velocity}]. Interestingly, it has also been proposed that driving infrared active phonons by ultrafast laser irradiation could induce superconductivity at temperatures much higher than the equilibrium critical one [\onlinecite{knap2015dynamical}]. 

Here, we revisit the polaron problem out of equilibrium when the electrons are periodically driven and show through explicit expressions how the binding energy and effective mass of the polaron can be controlled from the driving strength. To this purpose, we address the problem of noninteracting electrons that are rapidly driven and linearly coupled to vibrational modes in a one-dimensional crystal. Contrary to most of the nonequilibrium papers that we have mentioned so far and that deal with the real-time dynamics of an electron-phonon system, we rather focus on its stroboscopic dynamics, which is apprehended up to the third-order in the high-frequency expansion. This analytical approach provides a time-independent description of the problem in term of an effective Hamiltonian. In the absence of vibrational modes, it is well known that the Bloch band structure is simply renormalized by the time-periodic driving, which can result in the dynamical Wannier-Stark localization of electrons [\onlinecite{PhysRevB.34.3625}]. To our knowledge, this effect was first considered in Ref.\,[\onlinecite{Vonsovsky1939}]. In the presence of vibrational modes, we show that the driving actually modifies the electron-phonon interaction which becomes dynamically controllable when varying the driving strength. In order to be more specific, we focus on organic molecular crystals with electron-phonon interaction of Holstein type in equilibrium. Out of equilibrium, the driving additionally generates tunable nonlocal Peierls interactions and phonon-assisted hopping between distant neighbors. It turns out that both the phonon-assisted distant hopping and the renormalized nearest-neighbor tunneling can be dynamically suppressed when varying the driving strength. However, they cannot be suppressed simultaneously, meaning that the dynamical Wannier-Stark localization can no longer occur when the electrons are allowed to dissipate their energy on the vibrational modes of the crystal. Besides, we report the controllable nonequilibrium binding energy and effective mass of the polaron that the local and nonlocal electron-phonon interactions induce. This is achieved within both the weak- and strong-coupling regimes, since varying the driving strength enables the system to visit these two regimes dynamically.

While the high-frequency limit and simulations of lattice vibrations are already relevant in optical lattices of cold atomic gases [\onlinecite{lignier2007dynamical,eckardt2009exploring,struck2012tunable,greschner2014density,goldman2014periodically,PhysRevA.76.011605}], the explicit knowledge of the electron-phonon mechanisms we derive here in the third-order expansion allows the description of slower frequencies that become reasonable for solid state physics too, for example during multicycle laser irradiations in pump-probe experiments. The dynamical control allowed by the driving strength offers several opportunities among which the possibility to test weak- and strong-coupling polaron theories within a single material, or to understand a bit more the interplay between local and nonlocal electron-phonon interactions in organic crystals.

%%%
\section{Dynamical electron-phonon coupling}

\subsection{Time-periodic Hamiltonian}
When a homogeneous time-periodic electric field with magnitude $E_{0}$ and frequency $\Omega$ is driving noninteracting electrons in a one-dimensional crystal, it yields a vector potential that can be written as $A(t) = - E_{0} \sin (\Omega t) /\Omega$. The scalar potential is not relevant here for we consider the temporal gauge. Moreover Planck constant and the light celerity are set to unity, i.e. $\hbar=c=1$, and we chose the interatomic distance as unit of length. If the charge carriers are additionally coupled to vibrational modes, the system can generically be described by a time-periodic Hamiltonian of the form $H(t) = H_{e}(t) + H_{p} + H_{ep}$, with
\begin{align}\label{Time-Dependent Hamiltonian}
&H_{e}(t) =\sum_{k} \epsilon_{k}(t) \, c^{\dagger}_{k}c_{k} ~, ~~~ H_{p} =  \sum_{q} \omega_{q} \, b^{\dagger}_{q}b_{q} ~, \notag \\
&H_{ep} = \sum_{k,q} g_{q} \, c^{\dagger}_{k+q}c_{k}B_{q} ~.
\end{align}
According to Peierls substitution, the electronic dispersion relation is given by $\epsilon_{k}(t)=2\nu\cos( k+z\sin\Omega t)$, where $\nu$ refers to the nearest-neighbor hopping amplitude, $z=eE_{0}/\Omega$, and $e$ denotes the electron charge. In the model we are concerned with, the electrons are assumed to be linearly coupled to the atomic displacement operator $B_{q}=b^{\dagger}_{-q}+b_{q}$ through the coupling constant $g_{q}$, while $\omega_{q}$ defines the dispersion relation of phonons. No assumptions are made over these $q$-dependent functions for the moment.

%%%
\subsection{Third-order high-frequency description}
The dynamics of a quantum state $\phi (t)$ is then ruled by the time-dependent Schr\"odinger equation 
\begin{align}
i \, \partial_\tau \phi(\tau) =~\lambda \, H(\tau) \, \phi (\tau) ~,
\end{align} 
where $\tau=\Omega t$ and $\lambda=\delta E / \Omega$. Here $\delta E$ denotes a certain energy scale involved in the Hamiltonians of Eq.\,(\ref{Time-Dependent Hamiltonian}). Consequently, $\tau$ and $H(\tau)$ are dimensionless, though we still refer to them as time and Hamiltonian, respectively.

The high-frequency limit corresponds to $\lambda \ll 1$ or equivalently to $\delta E \ll \Omega$. If $\delta E$ is chosen as the largest characteristic energy scale met in Eq. (\ref{Time-Dependent Hamiltonian}), then there are no resonances with the driving which is said to be off-resonant. This limit can be apprehended through several analytical approaches among which Floquet-Magnus expansion, van Vleck and Brillouin-Wigner perturbation theories [\onlinecite{0305-4470-34-16-305,1367-2630-17-9-093039,PhysRevB.93.144307}]. Here we use a method which has been reported in Refs.\,[\onlinecite{Itin2}] and [\onlinecite{PhysRevLett.115.075301}]. It relies on the gauge transformation $\tilde{\phi} (\tau) = \exp\{-i\Delta(\tau)\}\,\phi (\tau)$, where $\Delta(\tau) = \sum_{n=1}^{+\infty} \Delta_{n}(\tau)\lambda^{n}$. Starting from the lowest order in $\lambda$, we iteratively build up operator $\Delta(\tau)$ under the constraint that $\Delta_{n}(\tau)$ is $2\pi$-periodic and averages at zero. The latter boundary condition ensures, similarly to van Vleck and Brillouin-Wigner approaches, that the perturbation theory does not depend on the arbitrary phase of the periodic driving [\onlinecite{PhysRevB.93.144307}]. By construction, this transformation is also required to remove the time-dependence of $H(\tau)$ in all orders in $\lambda$. So we end up with the effective Hamiltonian
\begin{align}\label{Effective Hamiltonian}
\tilde{H}=\lambda e^{i\Delta(t)} H(t) e^{-i\Delta(t)} -i e^{i\Delta(t)} \partial_{t} e^{-i\Delta(t)} 
\end{align}
that is time independent and also satisfies a Schr\"odinger-like equation:
\begin{align}
i\partial_\tau \tilde{\phi}(\tau) = \tilde H \tilde{\phi} (\tau) ~.
\end{align}
When assuming $\tilde{H}=\sum_{n=1}^{+\infty} \tilde{H}_{n}\lambda^{n}$ and restricting the high-frequency analysis to the third order in $\lambda$, Eq. (\ref{Effective Hamiltonian}) leads to
\begin{align}\label{Time-dependent Hamiltonians}
\tilde H_{1} &= H(\tau)-\partial_{\tau}\Delta_{1}(\tau) ~, \notag \\
\tilde H_{2} &= \frac{i}{2}[\Delta_{1}(\tau),H(\tau)]+\frac{i}{2}[\Delta_{1}(\tau),\tilde H_{1}]-\partial_{\tau}\Delta_{2}(\tau) \notag \\
\tilde H_{3} &= \frac{i}{2}[\Delta_{2}(\tau),H(\tau)]+\frac{i}{2} [\Delta_{1}(\tau),\tilde H_{2}] + \frac{i}{2} [\Delta_{2}(\tau),\tilde H_{1}] ~, \notag \\
&+ \frac{1}{12}[[\Delta_{1}(\tau),\partial_{t}\Delta_{1}(\tau)],\Delta_{1}(\tau)] - \partial_{\tau}\Delta_{3}(\tau)
~,
\end{align}
where the brackets refer to standard commutators. Since $\tilde H_{1}$, $\tilde H_{2}$ and $\tilde H_{3}$ have to be static by construction, they must be equal to their time average. Then taking the time average of the right-hand side terms in Eq.\,(\ref{Time-dependent Hamiltonians}) results in
\begin{align}\label{Time-independent Hamiltonians}
&\tilde H_{1} = H_{0} ~, ~~~~~~~~~~~~~~
\tilde H_{2} = -\frac{1}{2}\sum_{m\neq0}\frac{[H_{m},H_{-m}]}{m} ~, \\
&\tilde H_{3} = \frac{1}{2} \sum_{m\neq 0}\frac{[[H_{m},H_{0}],H_{-m}]}{m^{2}} + \frac{1}{3}\sum_{m\neq 0}\sum_{n\neq 0,m} \frac{[[H_{m},H_{n-m}],H_{-n}]}{mn} ~, \notag
\end{align}
where $H_{m}=\int_{-\pi}^{+\pi} \frac{d\tau}{2\pi}~ e^{im\tau} H(\tau)$. The first order simply refers to the time-averaged Hamiltonian because the electrons cannot follow the dynamics of the driving. Higher orders are commutation-based corrections that describe emissions and absorptions of virtual photons. As a result, the averaging method introduced above leads to time-independent effective Hamiltonians that describe the stroboscopic dynamics, whereas the evolution between two stroboscopic times is encoded into the operators $\Delta_{n}(\tau)$.

Importantly, the first and second orders of the high-frequency expansion are already realistic in systems such as ultracold atomic gases, for expample when shaking optical lattices with frequencies of a few $kHz$ [\onlinecite{lignier2007dynamical,eckardt2009exploring,struck2012tunable,greschner2014density}]. So the third-order description we address here may also be interesting to observe the effects of sub-$kHz$ frequencies in these systems. In solid state physics, however, rapidly driving electrons in the high-frequency limit faces several issues. On the one hand, the interesting effects predicted for noninteracting electrons such as dynamical localization and symmetry-protected topological phase transitions are based on the condition $J_{0}(z)=0$. For the first root of the 0-th order Bessel function this condition already requires a driving strength satisfying $eE_{0}\sim 2.4\,\Omega$. As we shall see later on, the high-frequency expansion usually relies on $2\nu \ll \Omega$ and is basically valid for laser frequencies of a few $eV$. Therefore, the condition $eE_{0}\sim 2.4\,\Omega$ involves even more energetic intensities that, additionally to be already challenging technically, are very likely to burn the crystal where the typical atomic binding energy is of the order of a few $eV$ per Angstrom too for covalent bonds. This issue is no longer a problem when dealing with interactions because the interesting physics due to corrections arises with $J_{m}(z)$, meaning with nonzero-th order Bessel functions. So they start playing a role as soon as the driving is turned on and there are already interesting effects for $eE_{0}<2.4\,\Omega$. Moreover we provide a high-frequency description up to the third-order, which is also expected to describe effects of slower driving frequencies and is \textit{a priori} more reasonable for solid state physics. As far as we shall be concerned, the hopping amplitude is about $0.1eV$ in organic molecular crystal like pentacene [\onlinecite{Li:2011nr},\,\onlinecite{Li:2013eu}], so the high-frequency effects we address further should be relevant for $eE_{0} \sim \Omega \sim 1\,eV$,  namely infrared light of $241.8\,THz$.
On the other hand, even if one can describe how electronic states are changed out of equilibrium, the question of how to reach a steady regime and populate the states in order to probe observables in solid states physics experiments is still under investigations [\onlinecite{seetharam2015controlled,canovi2016stroboscopic,mori2016rigorous}]. Here, we do not regard this latter issue. Instead, we rather address what kinds of electron-phonon interactions are induced by the off-resonant driving and how these interactions modify the equilibrium polaronic states.

%%%
\subsection{Time-independent effective Hamiltonian}
Now we are ready to apply the high-frequency approach introduced above to Hamiltonian $H(t)$ defined in Eq. (\ref{Time-Dependent Hamiltonian}). Its time Fourier transform consists of
\begin{align}
H_{m} &= \sum_{k} \epsilon_{k,m} \, c^{\dagger}_{k}c_{k} + \left( H_{p} + H_{ep} \right) \delta_{m,0} ~,
\end{align}
where $\epsilon_{k,m}=\int_{-\pi}^{+\pi} \frac{d\tau}{2\pi}~ e^{im\tau} \epsilon_{k}(\tau)$. In the absence of phonons, $H_{m}$ is a quadratic scalar operator, and $[c^{\dagger}_{k}c_{k}, c^{\dagger}_{k'}c_{k'}]=0$ is responsible for the cancellation of all commutators in Eq.\,(\ref{Time-independent Hamiltonians}). In this case, the stroboscopic dynamics is only described by the time-averaged Hamiltonian
\begin{align}
\tilde H_{1} = \sum_{k} \epsilon_{k,0}(z) \, c^{\dagger}_{k}c_{k} ~,
\end{align}
where $\epsilon_{k,m}(z)= 2\nu J_{m}(z) \cos(k)  / \delta E$ and $J_{m}$ is the $m$-th order Bessel function of the first kind. Thus, the off-resonant driving renormalizes the hopping amplitudes and is likely to localize the electrons for driving strengths that satisfy $J_{0}(z)=0$, which yields the so-called dynamical Wannier-Stark ladder in the density of states [\onlinecite{PhysRevB.34.3625}].

Such a renormalization of the electronic band structure suggests that, in the presence of interactions, the system may dynamically visit weak-, intermediate-, and strong-coupling regimes, as well as the one of strictly localized electrons. Moreover the interactions are time independent, so they only appear through Fourier component $H_{0}$. As the latter is not involved in the definition of $\tilde{H}_{2}$ in Eq.\,(\ref{Time-independent Hamiltonians}), there is no contribution at the second order of the high-frequency limit and $\tilde{H}_{2}=0$. The third order in $\lambda$, however, does depend on $H_{0}$ and leads to
\begin{align}\label{Effective Hamiltonian H3}
\tilde H_{3} &= \frac{1}{2}\sum_{m\neq0} \sum_{k,k'} \frac{\epsilon_{k,m} \epsilon_{k',-m}}{m^{2}} [ [ c^{\dagger}_{k}c_{k}, H_{ep} ], c^{\dagger}_{k'}c_{k'} ] ~.
%&= - \frac{1}{2} \sum_{k,q} \frac{g_{q}}{2\nu} ~ \eta_{k,q}(z) ~ c^{\dagger}_{k+q}c_{k}B_{q}  ~,
\end{align}
Consequently, the electron-phonon interaction, though time independent, is responsible for additional corrections to the effective Hamiltonian. In the case of the electron-phonon interaction, the effective Hamiltonian can by rewritten as follows
$\tilde{H} = \tilde{H}_{e} + \tilde{H}_{p} + \tilde{H}_{ep} + o(\lambda^{3})$, where
\begin{align}\label{Effective Holstein Hamiltonians}
&\tilde{H}_{e} = \sum_{k} 2\tilde{t}_{1}(z) \cos(k) \, c^{\dagger}_{k}c_{k} ~, ~~~ \tilde{H}_{p} =  \sum_{q} \tilde{\omega}_{q} b^{\dagger}_{q}b_{q} ~, \notag \\
&\tilde{H}_{ep} = \sum_{k,q} \gamma_{k,q}(z) ~ c^{\dagger}_{k+q}c_{k}B_{q} ~,
\end{align}
while $\tilde{t}_{1}(z) = \tilde{\nu} J_{0}(z)$, $\tilde{\nu}=\nu/\Omega$ and $\tilde{\omega}_{q}=\omega_{q}/\Omega$. The effective electron-phonon coupling $\gamma_{k,q}$ is specified in the next section. The reader may also find a detailed discussion about the role played by generic kinds of interactions in the high-frequency description in Ref. [\onlinecite{1367-2630-17-9-093039}].

\begin{figure}[t]
\centering
$\begin{array}{cc}
\includegraphics[trim = 26mm 0mm 15mm 0mm, clip, width=4.1cm]{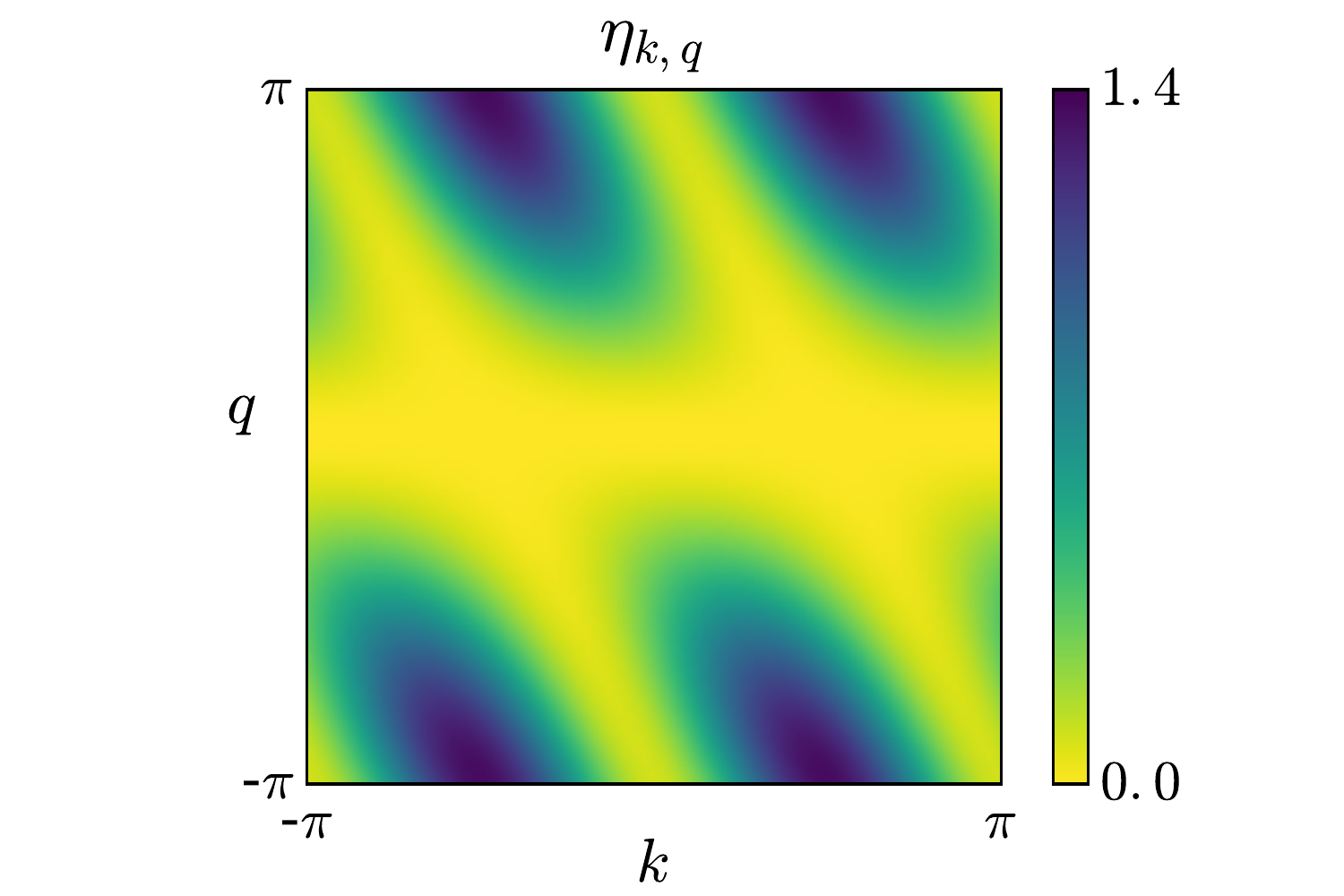} &
\includegraphics[trim = 26mm 0mm 15mm 0mm, clip, width=4.1cm]{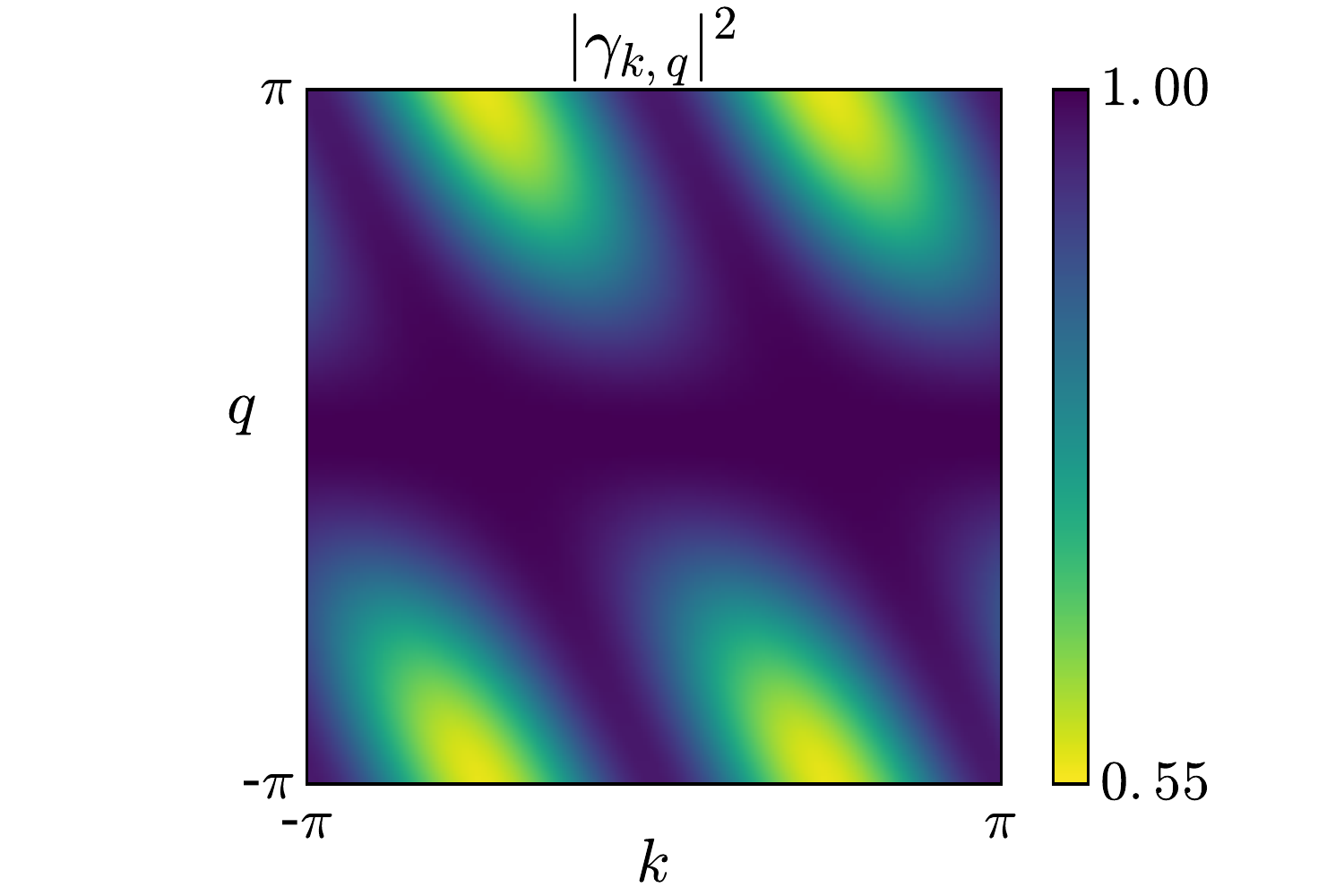}
\end{array}$
\caption{\small (Color online) Third-order correction $\eta_{k,q}$ (left) and effective electron-phonon coupling $\gamma_{k,q}$ in units of $g_{q}$ (right) for $\Omega=5\nu$ and $z=1.8$.}
\label{Photon-Renormalized Coupling}
\end{figure}

%%%
\subsection{Dynamical control of the electron-phonon coupling}
Whereas the phononic dispersion relation remains unchanged, the off-resonant driving renormalizes the electron-phonon interactions which, \textit{a priori}, becomes $k$-dependent out of equilibrium. This is characterized by the effective electron-phonon coupling
\begin{align}\label{Effective Coupling Gamma}
\gamma_{k,q}(z) = \tilde{g}_{q} \left( 1- \eta_{k,q}(z) \, \lambda^{2} \right) ~,
\end{align}
where $\eta_{k,q}$ arises from Eq\,(\ref{Effective Hamiltonian H3}) and appears as a second-order correction in $\lambda$ to the equilibrium electron-phonon coupling $\tilde{g}_{q}=g_{q}/\Omega$. It is given by
\begin{align}
\eta_{k,q}(z) = \sum_{m>0} \left( \frac{\bar{\epsilon}_{k+q,m}(z) - \bar{\epsilon}_{k,m}(z)}{m} \right)^{2} ~,
\end{align}
where $\bar{\epsilon}_{k,2n}=\epsilon_{k,2n}$ or $\bar{\epsilon}_{k,2n+1}=2\nu J_{2n+1}(z) \sin(k)  / \delta E$ for any integer $n$. This correction turns out to be positive for all strengths of the driving. As a result, the minus sign in Eq.\,(\ref{Effective Coupling Gamma}) suggests that it can only reduce the equilibrium electron-phonon coupling. The reader may find more details about the derivation of $\eta_{k,q}$ in Appendix. It is also straightforward to show that maxima of $\eta_{k,q}$ lye along the line $(k,0)$ in the $kq$-plane, whereas minima are located at $\pm(\pm \frac{\pi}{2}, \pi)$, in agreement with the map in Fig.\,\ref{Photon-Renormalized Coupling}. Thus, the effective electron-phonon coupling $|\gamma_{k,q}|^{2}$ favors the interactions between electrons and long-wavelength phonons $q\simeq 0$, as well as interactions with phonons of wavevectors $q\simeq -2k \pm \pi$. In this sense, the off-resonant driving acts as an interaction selector and can be regarded as a way to control the electron-phonon coupling in a dynamical and reversible way.

\begin{figure}[t]
\centering
$\begin{array}{c}
\includegraphics[trim = 5mm 0mm 10mm 0mm, clip, width=5cm]{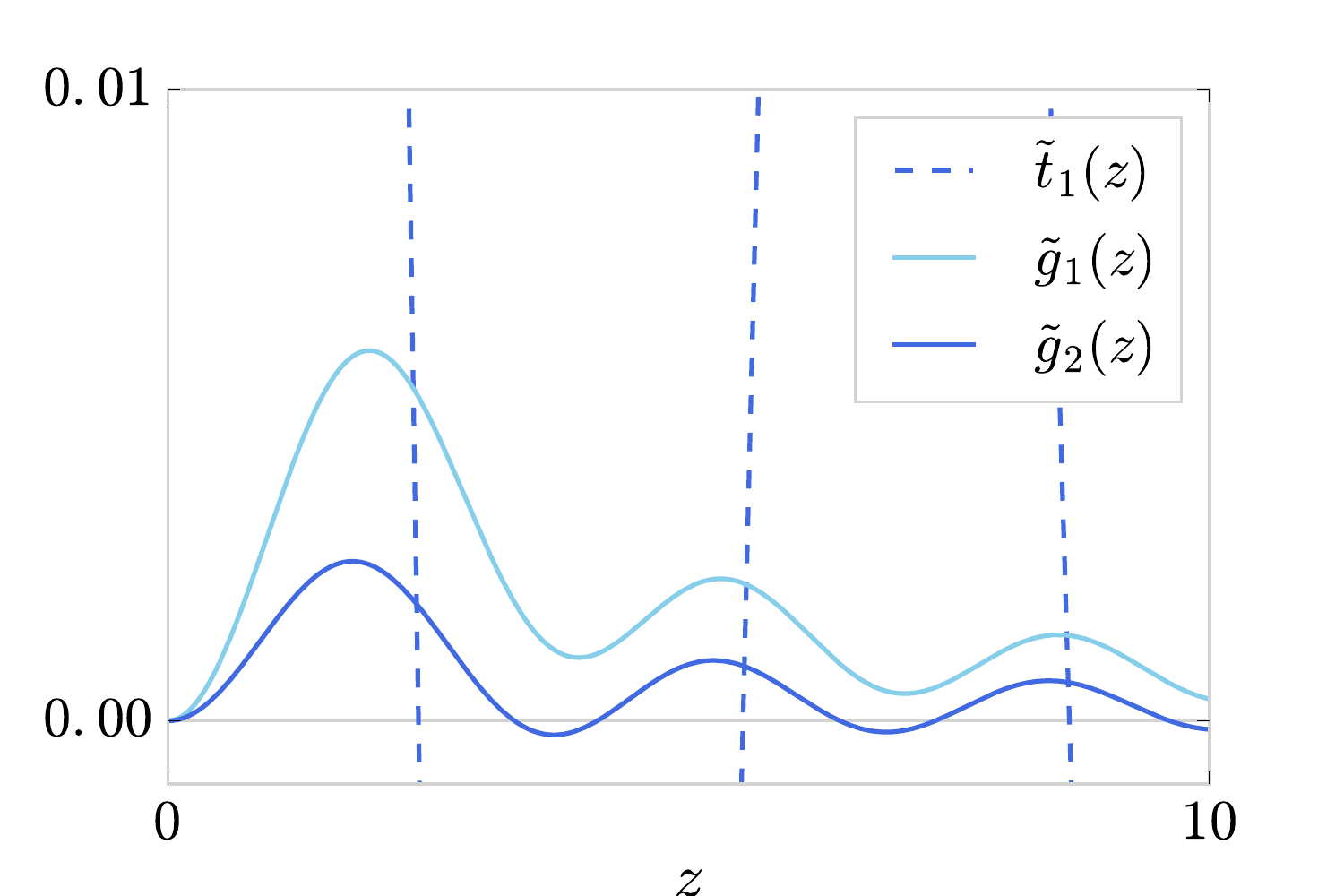}
\end{array}$
\caption{\small (Color online) Field-renormalized hopping and nonequilibrium corrections to the electron-phonon interaction as a function of the driving strength for $\Omega=5\nu$ and $g_{0}=\nu$.}
\label{Renormalized Coupling}
\end{figure}

It is also instructive to rephrase the effective electron-phonon Hamiltonian in terms of real-space coordinates. In order to clearly highlight the microscopic processes generated by the off-resonant driving, we now focus on a Hamiltonian that describes local electron-phonon interactions in equilibrium, meaning $g_{q}=g_{0}$. This kind of interactions is for example relevant in the context of polarons in organic molecular crystals, as reported by Hostein [\onlinecite{holstein1959studies}]. As detailed in Appendix, the effective electron-phonon Hamiltonian can be written in real space as
\begin{align}\label{Real Space Hep}
\tilde{H}_{ep} &= \tilde{g}_{0} \sum_{n} c^{\dagger}_{n}c_{n} \, B_{n} \\
&+ \tilde{g}_{1} (z) \sum_{n} c^{\dagger}_{n}c_{n}\left(B_{n-1}-2B_{n}+B_{n+1}\right) \notag \\
&+ \tilde{g}_{2}(z) \sum_{n} c^{\dagger}_{n}c_{n+2}\left(B_{n}-2B_{n+1}+B_{n+2}\right) + h.c. \notag
\end{align}
where the different electron-phonon couplings are defined by
\begin{align}\label{Coupling Definitions}
&\tilde{g}_{0} = \frac{g_{0}}{\Omega} ~, ~~~~~~~~ \tilde{g}_{1}(z) = \frac{1}{2} \frac{g_{0}}{\Omega} \left( \frac{2\nu}{\Omega} \right)^{2} \sum_{m>0}\frac{J_{m}^{2}(z)}{m^{2}} ~, \\
&\tilde{g}_{2}(z) = \frac{1}{4} \frac{g_{0}}{\Omega} \left( \frac{2\nu}{\Omega} \right)^{2} \sum_{m>0} \left( \frac{J_{2m-1}^{2}(z)}{(2m-1)^{2}} - \frac{J_{2m}^{2}(z)}{(2m)^{2}} \right) ~. \notag
\end{align}
Coupling $\tilde{g}_{0}$ comes from the time-averaged Hamiltonian $\tilde{H}_{1}$ and refers to Holstein local interactions as defined in equilibrium. Coupling $\tilde{g}_{1}$ is a nonequilibrium correction that simulates Peierls antisymmetric nonlocal interactions [\onlinecite{munn1985theory}], as introduced in the so-called SSH model to explain the formation of topological solitons in polyacetylene [\onlinecite{su1979solitons}]. Coupling $\tilde{g}_{2}$ is a nonequilibrium correction too. It describes phonon-assisted next-nearest-neighbor hopping processes. Both $\tilde{g}_{1}$ and $\tilde{g}_{2}$ refer to antisymmetric nonlocal interactions, which could already be known from the map $\gamma_{k,q}$ in Fig.\,\ref{Photon-Renormalized Coupling}, accordingly to the study of the symmetry effects of nonlocal electron-phonon interactions in Ref. [\onlinecite{Li:2011nr}]. Besides, $\tilde{g}_{1}$ and $\tilde{g}_{2}$ can both be controlled dynamically via the driving strength, as illustrated in Fig.\,(\ref{Renormalized Coupling}). Importantly, the phonon-assisted hopping processes can be turned off for some specific driving strengths. However, it cannot vanish simultaneously with the field-renormalized hopping $\tilde{t}_{1}$, which means that the noninteracting electrons can no longer experience the dynamical Wannier-Stark localization in the presence of lattice vibrations. It is worth mentioning that a similar conclusion holds when the electrons are driven by an electric field constant in time (instead of time-periodic). Indeed, the DC field leads to the Wannier-Stark localization (instead of dynamical Wannier-Stark localization) of the noninteracting electrons, but they get delocalized when they are coupled to lattice vibrations [\onlinecite{PhysRevB.88.035132}].

Besides, third-order corrections $\tilde{g}_{1}$ and $\tilde{g}_{2}$ scale with the factor $(2\nu / \Omega)^{2}$, regardless of the energy scale $\delta E$ we chose to define the small parameter $\lambda$ in the high-frequency expansion. As shown by Eq.\,(\ref{Effective Hamiltonian H3}), it is so because these corrections are defined from the square of harmonics of the electronic dispersion relation, whose characteristic energy scale corresponds to half the equilibrium bandwidth, namely $2\nu$. Of course, these corrections always remain small compared to Holstein coupling $\tilde{g}_{0}$. Nevertheless, they may compete the renormalized hopping processes when varying the driving strength $z$. Such a dynamical control, which should be suitable for multicycle laser pulse experiments and shaken optical lattices, may be useful for example to understand the role played by the nonlocal electron-phonon interactions in organic molecular semiconductors, where local Holstein interactions alone would not be sufficient to explain electronic transport [\onlinecite{zhao1994munn,Ciuchi:2011dn,Li:2011nr,Li:2013eu}].

%%%
\section{Effective Green functions}

\subsection{Perturbation theory along Schwinger-Keldysh contour}
\begin{figure}[t]
\centering
$\begin{array}{c}
\includegraphics[trim = 0mm 0mm 0mm 0mm, clip, width=4cm]{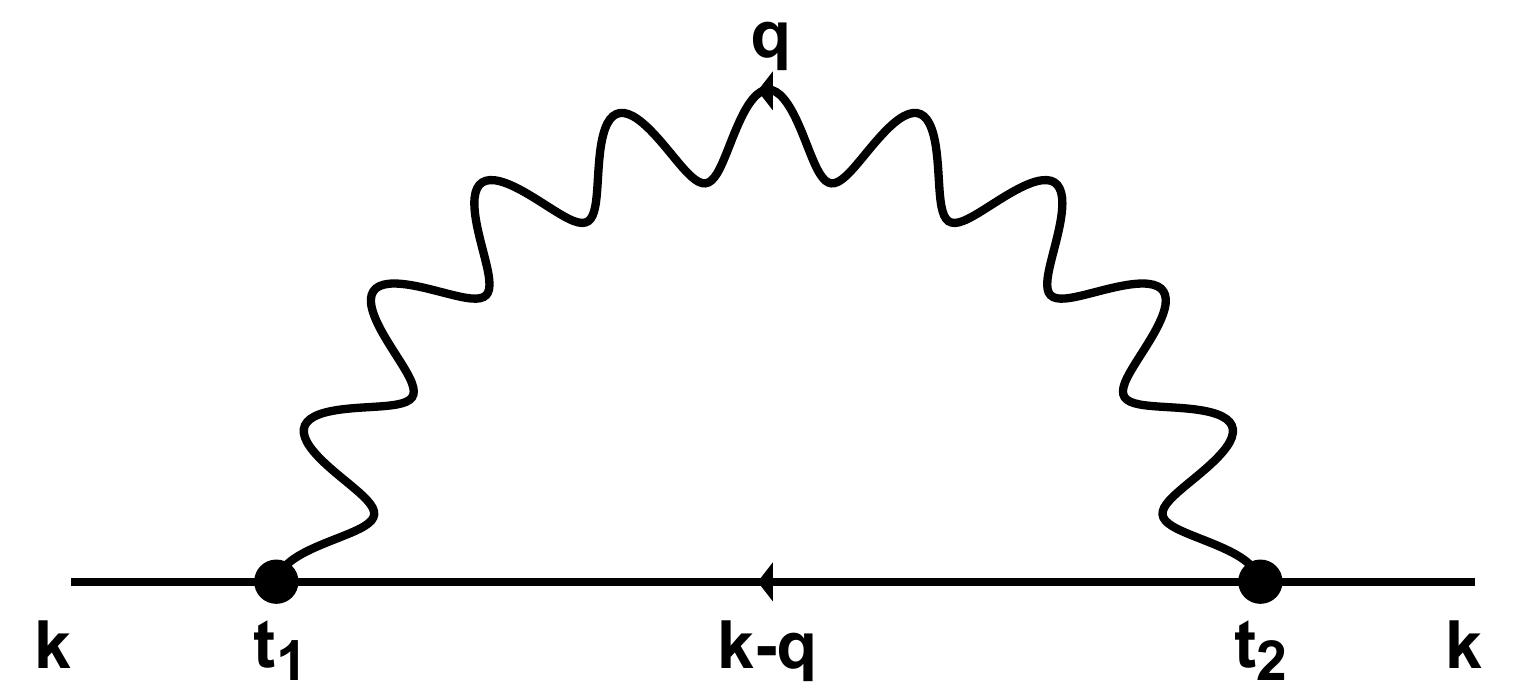}
\includegraphics[trim = 0mm 0mm 0mm 0mm, clip, width=4cm]{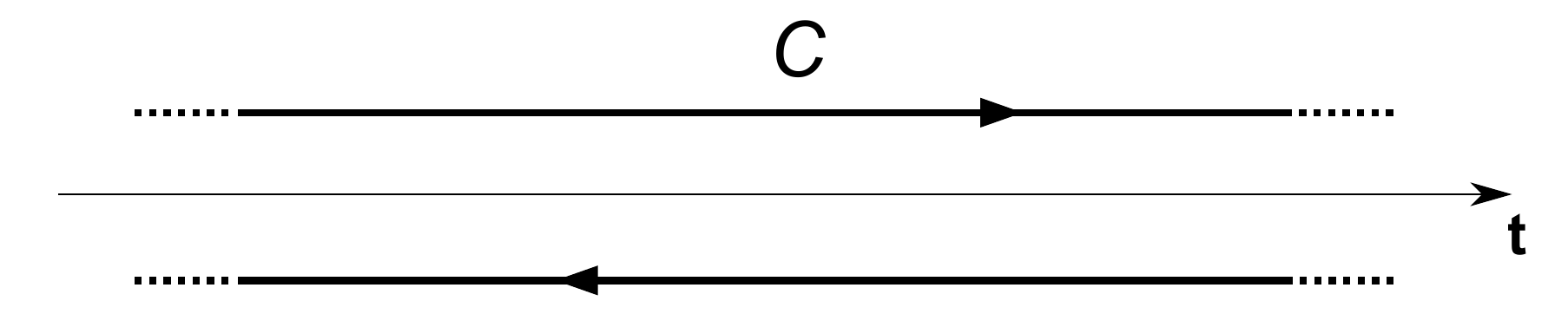}
\end{array}$
\caption{\small Diagrammatic representation of the electron-phonon interaction in a second-order perturbation theory (left) which is regarded here along Schwinger-Keldysh contour $C$ (right).}
\label{Diagram}
\end{figure}

Since the system is supposed to be in a nonequilibrium steady state, one can consider the time-dependent problem along the Schwinger-Keldysh contour $C$, as illustrated in Fig.\,\ref{Diagram}. In the interaction picture, the full Green function of the system can be written as a thermal average
\begin{align}\label{Full GF}
iG(k,t,t') = \big\langle {\cal{T}}_{C}e^{-i\int_{C}d\tau \sum_{k} V(k,\tau)}c_{k}^{~}(t)c_{k}^{\dagger}(t') \big\rangle_{0} ~,
\end{align}
where ${\cal{T}}_{C}$ denotes the time-ordering operator associated to the oriented contour $C$. The time evolution of operator $c_{k}(t)$ is ruled by the equation of motion based on time-dependent Hamiltonian $H_{e}(t)$ introduced in Eq.\,(\ref{Time-Dependent Hamiltonian}). Importantly the bracket index refers to the noninteracting density matrix of the system in equilibrium. This means that, first, we explicitly know the density matrix which is then given by $\rho_{0} = \frac{e^{-\beta H_{0}(-\infty)}}{\Tr [e^{-\beta H_{0}(-\infty)}]}$ and, second, we can take advantage of Wick theorem. The electron-phonon interaction is introduced as
\begin{align}
V(k,\tau) = \sum_{q} g_{q}~c_{k+q}^{\dagger}(\tau)c_{k}(\tau)B_{q}(\tau) ~.
\end{align}
In the framework of a perturbation theory, the first-order expansion in the electron-phonon coupling yields the thermal average of a single bosonic operator $B_{q}$ and therefore vanishes. Then the lowest-order contribution arises from the second-order, which leads to the following Green function 
\begin{align}\label{Contour 2nd Oder GF}
G^{(2)}(k,t,t') &= \frac{i}{2}\int_{C} dt_{1}dt_{2}\sum_{k_{1},k_{2}}\big\langle {\cal{T}}_{C}V(k_{1},t_{1})V(k_{2},t_{2})c_{k}^{~}(t)c_{k}^{\dagger}(t') \big\rangle_{0} \notag \\
&= \int_{C} dt_{1}dt_{2}~ G^{(0)}(k,t,t_{1}) \Sigma^{(2)}(k,t_{1},t_{2}) G^{(0)}(k,t_{2},t') ~,
\end{align}
The bare electron and phonon Green functions are respectively defined as $G^{(0)}(k,t,t') = \big\langle {\cal{T}}_{C}~ c_{k}(t)c_{k}^{\dagger}(t') \big\rangle_{0}$ and $D^{(0)}(q, t,t') = \big\langle {\cal{T}}_{C}~ B_{q}(t)B_{q}^{\dagger}(t') \big\rangle_{0}$. It corresponds to the Fock-like diagram illustrated in Fig.\,\ref{Diagram}. This is the single non-vanishing second-order contribution. It describes the emission of a phonon with momentum $q$ at $t_{2}$ and its subsequent absorption at $t_{1}$. The self-energy associated to this single-phonon process is
\begin{align}
\Sigma^{(2)}(k,t_{1},t_{2}) = i \int_{BZ} dq~ g_{q}^{2}~ G^{(0)}(k+q,t_{1},t_{2})~ D^{(0)}(q,t_{1},t_{2}) ~.
\end{align}
Considering that any time variable can be located either along the forward branch or along the backward one of contour $C$, it is then possible to rephrase this equation in terms of 2$\times$2 matrices. In Keldysh basis, the second-order Green function can be rewritten as
\begin{align}
G^{(2)}(t,t') &= \int \int dt_{1} dt_{2}~ G^{(0)}(t,t_{1}) \, \Sigma^{(2)}(t_{1},t_{2}) \, G^{(0)}(t_{2},t') ~,
\end{align}
where momentum $k$ have been omitted for more clearness, integral $\int$ runs from $t=-\infty$ up to $t=+\infty$ and
\begin{align}
G^{(0)}
&=
\left( \begin{array}{cc} 
G_{R}^{(0)} & G_{K}^{(0)} \\
0 & G_{A}^{(0)} \\
\end{array} \right) ~,~
D^{(0)}
=
\left( \begin{array}{cc} 
D_{R}^{(0)} & D_{K}^{(0)} \\
0 & D_{A}^{(0)} \\
\end{array} \right) ~,\\
\Sigma^{(2)} 
&=
\left( \begin{array}{cc} 
\Sigma_{R}^{(2)} & \Sigma_{K}^{(2)} \\
0 & \Sigma_{A}^{(2)} \\
\end{array} \right) ~. \notag
\end{align}
The indices $R$, $K$ and $A$ respectively label the retarded, Keldysh and advanced Green functions.

The retarded component of the self-energy in Keldysh formalism is
\begin{align}
\Sigma^{(2)}_{R}(k)= \frac{i}{2} \int_{BZ}dq \left[G_{R}^{0}(k+q) \, D_{K}^{0}(q) + G_{K}^{0}(k+q) \, D_{R}^{0}(q) \right] ~,
\end{align}
where the two time variables have been omitted for more clearness. Because the system is out of equilibrium, the two time variables of Green functions are independent. Then it is convenient to rephrase them in terms of the relative time $t=t_{1}-t_{2}$ and the averaged time $T=(t_{1}+t_{2})/2$ [\onlinecite{wigner1932quantum}]. This can be compared to the equilibrium situation, where Green functions only depend on the relative time, whose conjugate variable is the frequency $\omega$. The Fourier transform of the retarded and Keldysh Green functions, with respect to the relative time, leads to the following expression for the self-energy
\begin{align}
\Sigma^{(2)}_{R}(k,\omega,T) = \int_{BZ}dq~ g^{2}_{q}~ \Big \{ [N_{q}+n_{k+q}]G_{R}^{0}(k+q,\omega+\omega_{q},T)& \notag \\
+[N_{q}+1-n_{k+q}]G_{R}^{0}(k+q,\omega-\omega_{q},T)& \Big \} ~.
\end{align}
The functions $N_{q}$ and $n_{k+q}$ respectively denote the Bose-Einstein and Fermi-Dirac distributions, meaning the distributions for identical particles when the system was in equilibrium at time $t=-\infty$.

%%%
\subsection{Perturbation theory for effective Green functions}
The nonequilibrium perturbation theory along Schwinger-Keldysh contour refers to Green functions based on time-periodic Hamiltonian (\ref{Time-Dependent Hamiltonian}). At present we show that we can equivalently define effective Green functions based on time-independent effective Hamiltonian (\ref{Effective Holstein Hamiltonians}) that describes the system in the high-frequency limit. We can start from the equation of motion
\begin{align}
\left[ i \partial_{\tau} - \lambda H(\tau) \right] G(\tau, \tau') &= \delta(\tau, \tau')
\end{align}
and straightforwardly show that the gauge transformation introduced earlier to define the effective Hamiltonian leads to
\begin{align}
\left[ i\partial_{\tau} + \tilde{H} \right] \tilde{G}(\tau'-\tau) &= \delta(\tau, \tau') ~,
\end{align}
where we refer to $\tilde{G}(\tau'-\tau) = e^{i\Delta(\tau)}\,G(\tau, \tau')\,e^{-i\Delta(\tau')}$ as effective Green function. This is a one-time-argument function that describes a system invariant by time translation. Consequently, two stroboscopic times characterized by an integer $n$ such that $\tau'-\tau = n \, 2\pi$, along with the $2\pi$-periodicity of $\Delta(\tau)$, result in
\begin{align}
\Tr \tilde{G}(\tau'-\tau) = \Tr G(\tau, \tau') ~.
\end{align}
Observables such as the density of states are then equal in both descriptions. As far as we are concerned, the single-orbital electronic Green functions are scalars and then equal each other for stroboscopic times.

Now that we have introduced the notion of effective Green function in the high-frequency limit, we are ready to revisit the perturbation theory. The multiplicative structure of the Dyson equation is responsible for
\begin{widetext}
\begin{align} 
G(\tau, \tau') &= G^{0}(\tau, \tau') 
+ \int\int d\tau_{1} d\tau_{2} G^{0}(\tau, \tau_{1})\,\Sigma(\tau_{1},\tau_{2}) \, G^{0}(\tau_{2}, \tau')
+ ... \notag \\
&= e^{-i\Delta(\tau)} \, \tilde{G}^{0}(\tau'-\tau) \, e^{i\Delta(\tau')} 
+ e^{-i\Delta(\tau)} \, \int\int d\tau_{1} d\tau_{2} \tilde{G}^{0}(\tau_{1}-\tau)\,\tilde{\Sigma}(\tau_{2}-\tau_{1}) \,\tilde{G}^{0}(\tau'-\tau_{2}) \, e^{i\Delta(\tau')}
+ ... 
\end{align}
\end{widetext}
where $\tilde{\Sigma}(\tau'-\tau) = e^{i\Delta(\tau)} \, \Sigma(\tau',\tau) \, e^{-i\Delta(\tau')}$ defines the effective self-energy. As a result, there is a one to one correspondence at in all orders in the perturbation theory between the $n$-th order of the time-periodic problem and the $n$-th order of the time-independent effective problem. However, the interaction vertex $g$ the self-enerfy $\Sigma(\tau_{1},\tau_{2})$ relies on is renormalized in the effective description, meaning that $\tilde{\Sigma}(\tau_{2}-\tau_{1})$ refers to an effective interaction vertex $\tilde{g}$. In other words, the local-in-time gauge transformation $e^{i\Delta(\tau)}$ enables us to regard the time-evolution of the initial time-periodic system in terms of the evolution of an effective time-independent one with a renormalized band structure and renormalized interactions. This greatly simplifies the problem since we can simply use the standard rules for equilibrium Green functions.

For example, the second-order perturbation theory leads to the following retarded component for the effective self-energy:
\begin{widetext}
\begin{align}\label{Self-Energy}
\tilde{\Sigma}^{(2)}_{R}(k,\tilde{\omega}) &= \int_{BZ}dq \, \gamma_{k,q}\gamma_{k+q,-q} \left( \frac{N_{0} + n_{q+k} }{\tilde{\omega} + \tilde{\omega}_{0} - \epsilon_{k+q,0} + i\delta} +\frac{N_{0}+1-n_{q+k}}{\tilde{\omega} - \tilde{\omega}_{0} - \epsilon_{k+q,0} + i\delta} \right) ~,
\end{align}
\end{widetext}
where $N_{0}$ denotes the equilibrium distribution function of dispersionless phonons and $\delta$ is the inverse of the quasiparticle lifetime which is introduced in the definition of the bare Green function. The first term proportional to $N_{0}$ describes the absorbtion of a phonon, whereas the second term, which is proportional to $N_{0}+1$ and does not vanish even at zero temperature, corresponds to the emission of phonons by the electrons. Besides, the renormalized coupling preserves the Hermitian structure of the effective electron-phonon Hamiltonian and satisfies
\begin{align}
\gamma_{k,q}\gamma_{k+q,-q} &= |\gamma_{k,q}|^{2} \\
&= \tilde{g}_{0}^{2}(1-2\eta_{k,q}\lambda^{2}) + o (\lambda^{3}) ~, \notag
\end{align}
We remind the reader of the map $|\gamma_{k,q}|^{2}$ that has already been introduced in Fig.\,\ref{Photon-Renormalized Coupling}.

%%%
\section{Weak-coupling regime}

%%%
\subsection{Single electron properties}
Because the off-resonant driving renormalizes the electronic bandwidth, it enables the system to visit weak- and strong-couling regimes in a dynamical way. Here, we begin with the description of the weak-coupling regime, which corresponds to driving strengths $z$ that satisfy $\tilde{g}_{0} \ll |\tilde{t}_{1}(z)|$. Moreover, we consider that Eq.\,(\ref{Self-Energy}) does not depend on the fermionic statistic for we consider a single electron in the band, as assumed in Fr\"ohlich polaron problem [\onlinecite{frohlich1950xx,frohlich1952interaction,feynman1955slow}]. Within Holstein description of organic molecular crystals [\onlinecite{holstein1959studies}], an electron that hops onto a molecule excites a vibrational mode which subsequently relaxes after the electron moves away. The molecular displacement the electron induces along its motion results in a surrounding phonon cloud, which changes the electron energy and effective mass. This electron dressed by the lattice polarization is referred to as polaron. In the presence of off-resonant driving, one naturally expects third-order corrections $\tilde{g}_{1}$ and $\tilde{g}_{2}$ in Hamiltonian (\ref{Real Space Hep}) to modify the equilibrium polaronic properties. This is the purpose of the subsequent lines.

\subsubsection{Generic case}
\begin{figure}[t]
\centering
$\begin{array}{c}
\includegraphics[trim = 0mm 0mm 0mm 0mm, clip, width=6cm]{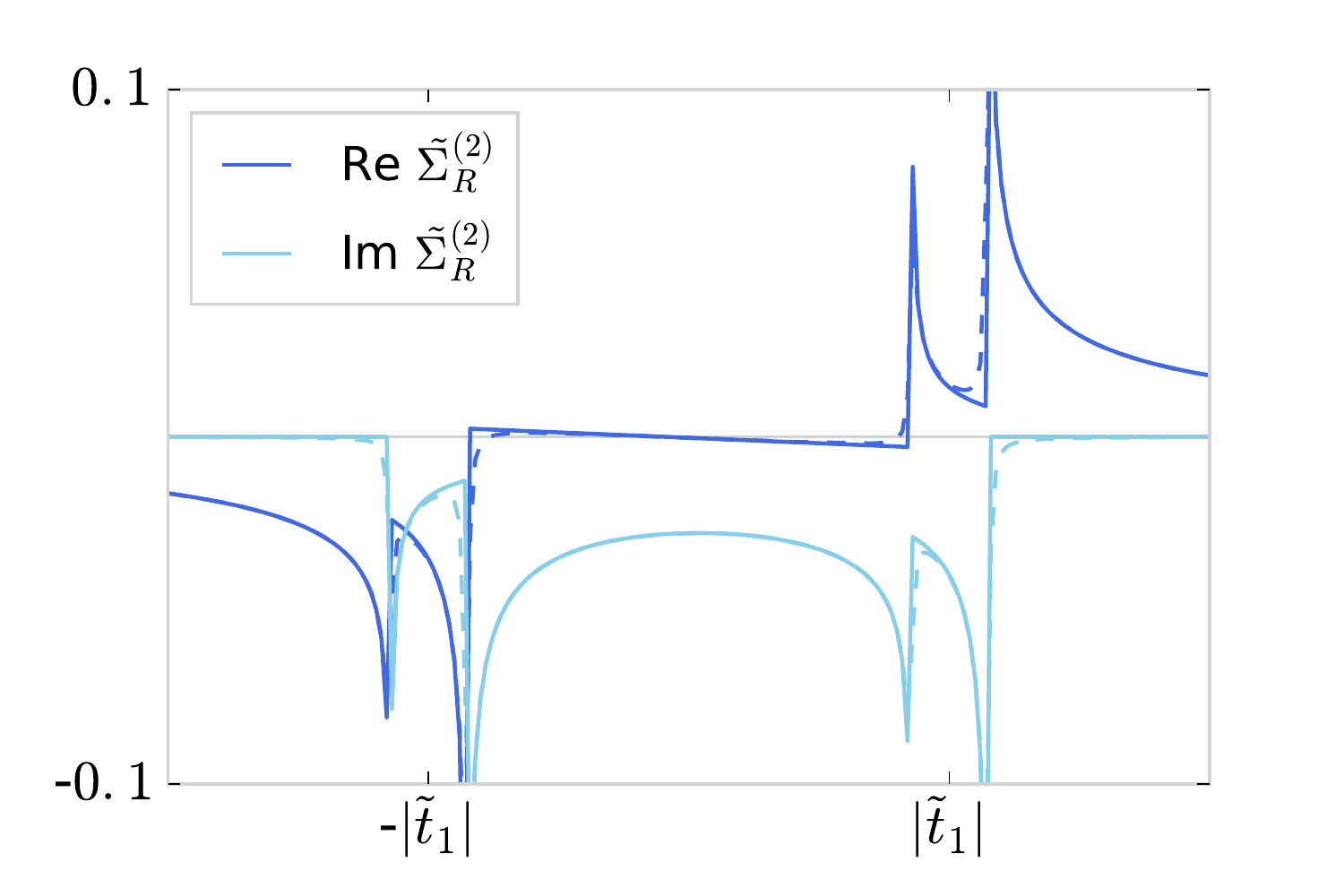}
\end{array}$
\caption{\small (Color online) Real and imaginary parts of the retarded component of the effective self-energy for a single electron at room temperature. Analytics (full lines) is compared to numerics (dashed lines) for $\Omega = 5\nu$, $\omega_{0}=0.1\nu$, $g_{0}=0.2\nu$, $z=1.8$, $\delta=0.01$ and $k=0$.}
\label{Effective SelfEnergy}
\end{figure}

First of all, it can be noticed that the retarded component of the effective self-energy in Eq.\,(\ref{Self-Energy}) is a complex function whose real and imaginary part can be known analytically and exactly for arbitrary parameters. Its expression is derived in Appendix but, because it is rather cumbersome, we do not present it in the main text. Instead, we present its real and imaginary parts in Fig.\,\ref{Effective SelfEnergy}, when there is a single electron in the band that is linearly coupled to vibrational modes at room temperature, i.e. $k_{B}T = 25 \,meV$. In this case, the electron is allowed to emit and absorb phonons. This yields two emission and two absorption peaks that are located at $|\tilde{\omega}-\tilde{\omega}_{0}|=2|\tilde{t}_{1}|$ and $|\tilde{\omega}+\tilde{\omega}_{0}|=2|\tilde{t}_{1}|$, respectively. Fig.\,\ref{Effective SelfEnergy} also compares our analytical evaluation of the effective self-energy to its numerical computation obtained from Eq.\,(\ref{Self-Energy}). They both exhibit the same behavior, the little error in between the full and dashed lines being due to the finite quasiparticle lifetime $1/\delta$ that is required to perform integral (\ref{Self-Energy}) numerically.

In order to get some more physical insight into this self-energy, we now focus on two peculiar situations, namely the adiabatic and non-adiabatic cases.

\subsubsection{Non-adiabatic limit $|\tilde{t}_{1}| \ll \tilde{\omega}_{0}$}
The non-adiabatic limit $|\tilde{t}_{1}| \ll \tilde{\omega}_{0}$ refers to a situation in which the electron tunneling is much slower than the vibrations of molecules. In the limit of small $k$, the retarded component of the effective self-energy introduced in Eq.\,(\ref{Self-Energy}) leads to the following polaronic dispersion relation
\begin{align}
\tilde{\xi}_{k} &= \epsilon_{k,0} + \Real \tilde{\Sigma}^{(2)}_{R}(k,\tilde{\xi}_{k}) \notag \\
&\simeq -\tilde{\Delta} + \frac{1}{1+ (2N_{0}+1) \frac{\tilde{\Delta}}{\tilde{\omega}_{0}}} \, \frac{k^{2}}{2\tilde{m}} ~.
\end{align}
This expression, which is derived in Appendix, looks like the one obtained at zero temperature in equilibrium [\onlinecite{klamt1988tight},\,\onlinecite{barivsic2008phase}]. However, the electron mass $\tilde{m}$ takes into account the flattening of the noninteracting electron band due to the time-periodic driving. So it is a function of the driving strength that is defined as
\begin{align}
\tilde{m}(z) = \frac{1}{\tilde{t}_{1}(z)} ~.
\end{align}
Moreover, the polaron binding energy is corrected by the electron-phonon couplings induced out of equilibrium. It is also a function of the driving strength and satisfies
\begin{align}\label{Binding Energy NonAdiabatic}
\tilde{\Delta}(z) = \frac{\tilde{g}_{0}^{2} - 4\tilde{g}_{0}\tilde{g}_{1}(z)+4\tilde{g}_{0}\tilde{g}_{2}(z)}{\tilde{\omega}_{0}} ~.
\end{align}
Finally the polaron mass $\tilde{m}^{*}$ depends on the phonon temperature and driving strength as
\begin{align}
\tilde{m}^{*}(z) = \left[ 1+ (2N_{0}+1)\frac{\tilde{\Delta}(z)}{\tilde{\omega}_{0}} \right] \tilde{m}(z) ~.
\end{align}
When the off-resonant driving is turned off, the binding energy reduces to $\tilde{\Delta}(0) = \tilde{g}_{0}^{2}/\tilde{\omega}_{0}$ and the expressions above are in agreement with the polaron behavior in equilibrium [\onlinecite{klamt1988tight},\,\onlinecite{barivsic2008phase}].

\subsubsection{Adiabatic limit $\tilde{\omega}_{0} \ll |\tilde{t}_{1}|$}
The adiabatic limit $\tilde{\omega}_{0}\ll |\tilde{t}_{1}|$ corresponds to the case of an electron hopping that is much faster than the vibrations of the lattice. This limit is for instance relevant when the electron-phonon coupling is weak ($\tilde{g}_{0} \ll |\tilde{t}_{1}|$) in organic molecular crystals like pentacene where $\tilde{g}_{0} \sim \tilde{\omega}_{0}$ [\onlinecite{Li:2011nr},\,\onlinecite{Li:2013eu}].

When $-2|\tilde{t}_{1}| + \tilde{\omega}_{0} < \tilde{\omega} < 2|\tilde{t}_{1}| - \tilde{\omega}_{0}$, we obtain from Eq.\,(\ref{Self-Energy}) a simple expression for the polaronic dispersion relation, namely
\begin{align}
\tilde{\xi}_{k} &=  \tilde{\Delta} + \frac{\tilde{m}}{\tilde{m}^{*}} \, \epsilon_{k,0} ~.
\end{align}
Note that this expression holds for all values of $k$ within the Brillouin zone, so it characterizes a whole polaron band. The onsite energy felt by the polaron is
\begin{align}
\tilde{\Delta}(z) = 2\frac{\tilde{g}_{0}\tilde{g}_{2}(z)}{\tilde{t}_{1}^{2}(z)} \, \tilde{\omega}_{0}
\end{align}
and its effective mass is defined by
\begin{align}
\tilde{m}^{*}(z) = \left[ 1+ (2N_{0}+1) \frac{\tilde{g}_{0}\tilde{g}_{1}(z)}{\tilde{t}_{1}^{2}}\right] \tilde{m}(z) ~.
\end{align}
Contrary to the non-adiabatic case, the onsite energy $\tilde{\Delta}$ can dynamically change signs as a function of the driving strength. Therefore, it does not necessarily refer to a binding energy since, when $\tilde{\Delta} >0$, the polaron feels a repulsive potential on each lattice site. The effective mass, however, is always heavier than it is in equilibrium because, first, the driving flattens the curvature of the electronic band and, second, the electron drags the phonon cloud along its motion. It is also worth mentioning that the onsite energy felt by the polaron and its effective mass both vanish in equilibrium and consist of purely out-of-equilibrium polaronic effects.

Moreover, the polaron energy $\tilde{\xi}_{k}$ is larger than the phonon frequency $\tilde{\omega}_{0}$. Thus, the polaron can also emit a phonon, even at zero temperature when $N_{0}=0$, which yields a nonzero imaginary part to the self-energy. The zeroth order in the adiabatic limit $\tilde{\omega}_{0} \ll |\tilde{t}_{1}|$ leads to a  scattering time $\tilde{\tau}$ that satisfies
\begin{align}
\frac{1}{\tilde{\tau}(k,\tilde{\omega})} &= - \Imag \tilde{\Sigma}^{(2)}_{R}(k,\tilde{\omega})\notag \\
&=
\frac{2N_{0}+1}{\sqrt{4\tilde{t}_{1}^{2}-\tilde{\omega}^{2}}}
\Bigg[
\tilde{g}_{0}^{2} - \tilde{g}_{0}\tilde{g}_{1} \left( 4 - \frac{\epsilon_{k,0}}{\tilde{t}_{1}} \frac{\tilde{\omega}}{\tilde{t}_{1}} \right) \notag \\
&- \tilde{g}_{0}\tilde{g}_{2} \left( 4 - 2\frac{\epsilon_{2k,0}}{\tilde{t}_{1}} + 2\frac{\epsilon_{k,0}}{\tilde{t}_{1}} \frac{\tilde{\omega}}{\tilde{t}_{1}} - 2 \frac{\tilde{\omega}^{2}}{\tilde{t}_{1}^{2}} \right)
\Bigg] ~.
\end{align}
The polaron lifetime is already finite in equilibrium but the nonequilibrium corrections make it $k$-dependent.

\begin{figure}[t]
\centering
$\begin{array}{cc}
\includegraphics[trim = 17mm 0mm 25mm 0mm, clip, width=4.2cm]{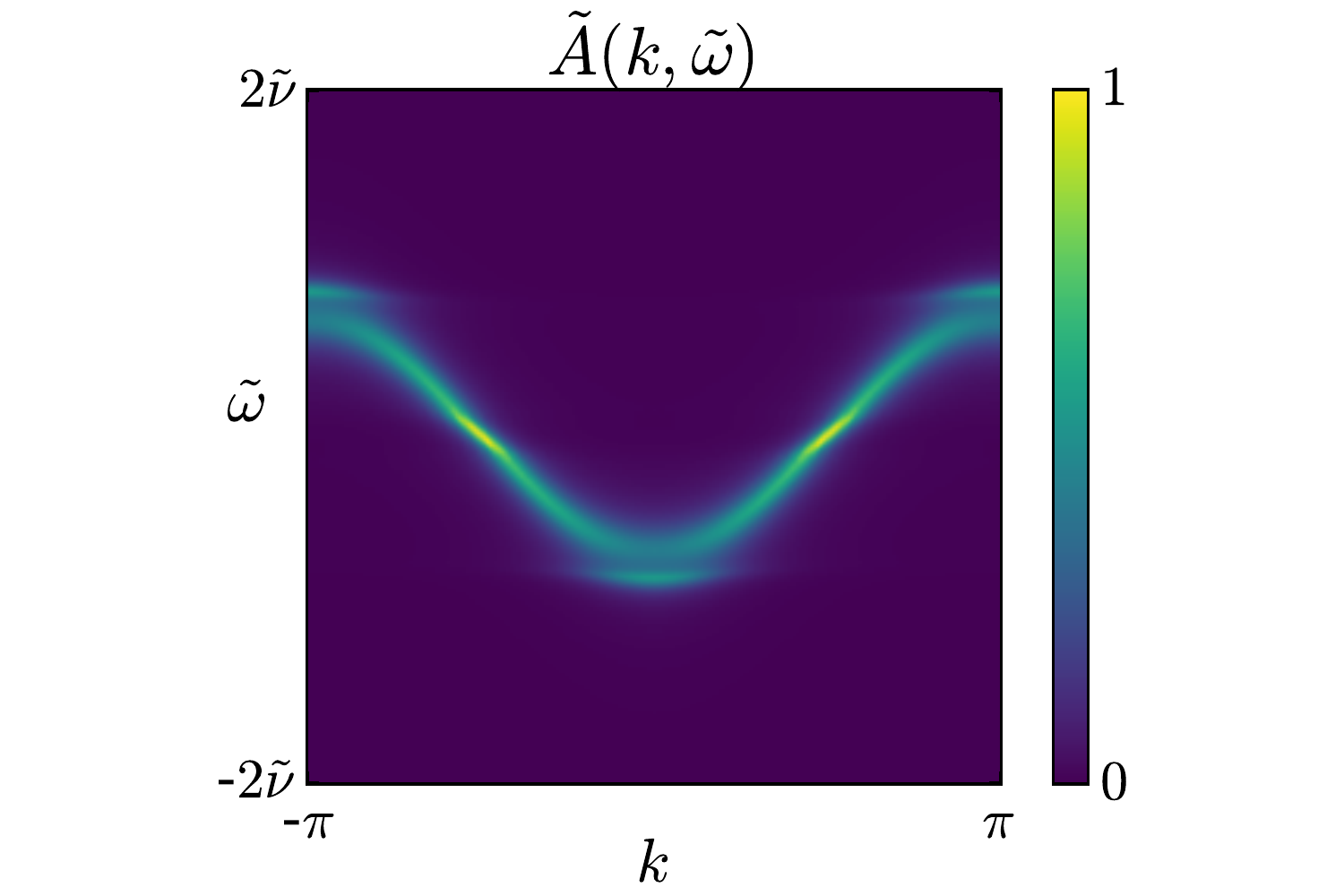} &
\includegraphics[trim = 17mm 0mm 25mm 0mm, clip, width=4.2cm]{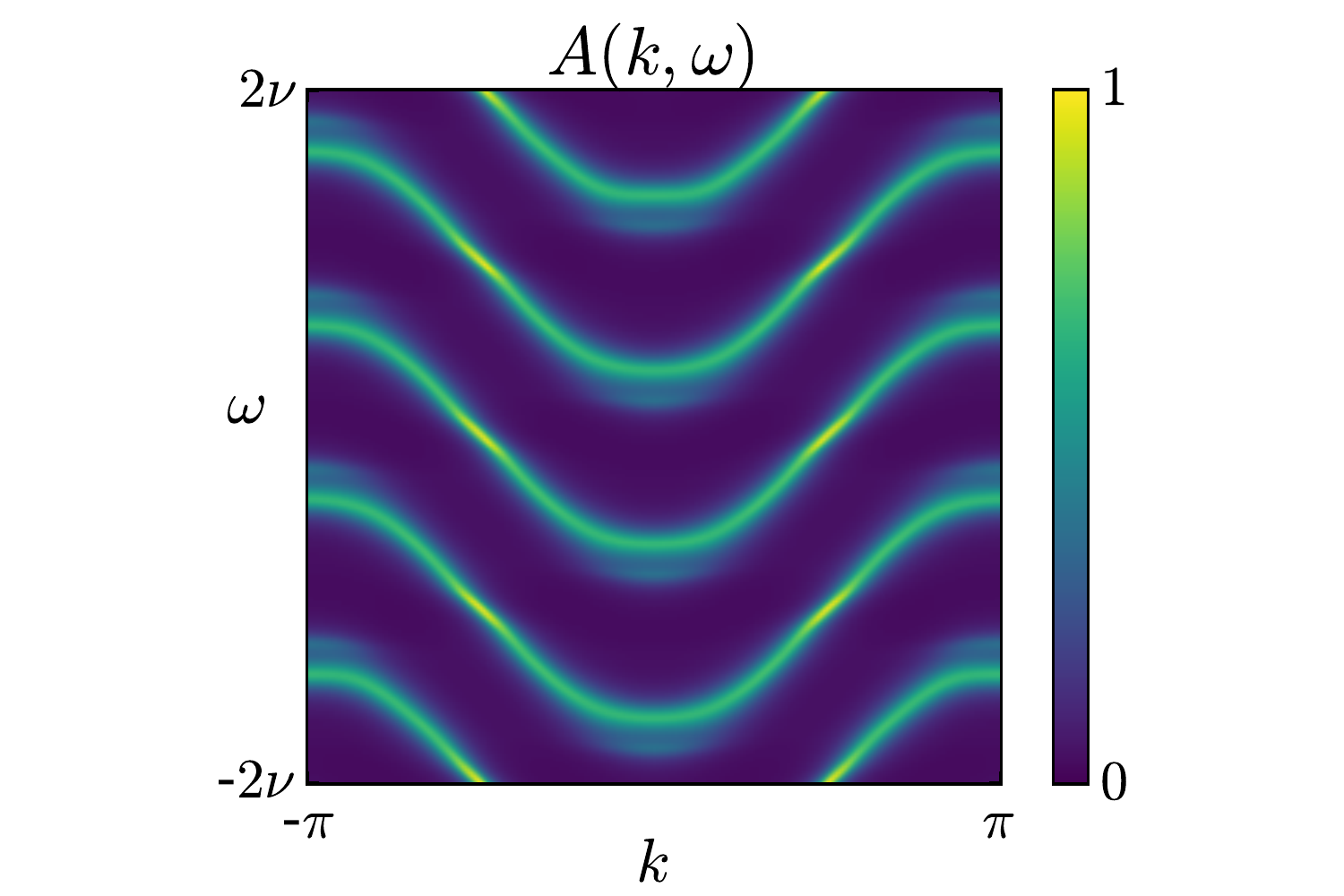} \\
\end{array}$
\caption{\small (Color online) Effective and Floquet spectral functions for $\Omega = 5\nu$ (left) and $\Omega = \nu$ (right), respectively. Both spectral functions have been computed for zero temperature with the following parameters: $\omega_{0}=0.1\nu$, $ g_{0}=0.2\nu$, $z=1.8$, and $\delta=0.01$.}
\label{Spectral Function}
\end{figure}

When $-2\tilde{t}_{1} - \tilde{\omega}_{0} < \tilde{\omega} < - 2\tilde{t}_{1} + \tilde{\omega}_{0}$, we can also determine the polaron properties for energies in the vicinity of $-2\tilde{t}_{1}$. The reader may refer to Appendix for more details. Such energies are associated to the bottom of the equilibrium electron band, for we consider, without loss of generality, that $\tilde{t}_{1}(z)>0$. Then Eq.\,(\ref{Self-Energy}) leads to the following polaronic dispersion relation in the limit of small $k$
\begin{align}
\tilde{\xi}_{k} &\simeq -\tilde{\Delta} + \frac{1}{1+ \frac{\tilde{\Delta}}{2\tilde{\omega}_{0}}} \frac{k^{2}}{2\tilde{m}} ~.
\end{align}
The onsite energy felt by the polaron is now negative and again defines a binding energy with
\begin{align}
\tilde{\Delta}(z) = (N_{0}+1) \, \frac{\tilde{g}_{0}^{2} - 8\tilde{g}_{0}\tilde{g}_{1}(z)}{\sqrt{4\tilde{\omega}_{0}\tilde{t}_{1}(z)}} ~.
\end{align}
Note that this is a function of the phonon temperature too. Besides the effective mass of the polaron is given by
\begin{align}
\tilde{m}^{*}(z) = \left[ 1+ \frac{\tilde{\Delta}(z)}{2\tilde{\omega}_{0}} \right] \tilde{m}(z) ~.
\end{align}
Again we can check that, when the off-resonant driving is turned off, the binding energy reduces to $\tilde{\Delta} = \tilde{g}_{0}^{2}/\sqrt{4\tilde{\nu}\tilde{\omega}_{0}}$, so that the expressions above yield the same results as the equilibrium ones [\onlinecite{klamt1988tight},\,\onlinecite{barivsic2008phase}].

%%%
\subsection{Effective and Floquet spectral functions}

The retarded component of the effective self-energy introduced in Eq.\,(\ref{Self-Energy}) leads to the effective spectral function
\begin{align}
\tilde{A}(k,\tilde{\omega}) \simeq -\frac{1}{\pi} \Imag \left[ \tilde{G}^{0}_{R}(k,\tilde{\omega}) - \tilde{\Sigma}^{(2)}_{R}(k,\tilde{\omega}) \right]^{-1} ~.
\end{align}
Importantly, the effective spectral function is a gauge invariant quantity, since it has been introduced in the context of the stroboscopic dynamics and, therefore, it is not affected by the momentum shift required to make Green functions gauge invariant out of equilibrium [\onlinecite{boulware1966gauge,davies1988narrow,aoki2014nonequilibrium}]. Note moreover that Keldysh approach relies on the equilibrium Fermi-Dirac distribution, since it assumes that the system was in equilibrium at time $\tau=-\infty$. This is the reason why the equilibrium distribution function appears in the expression of the effective self-energy in Eq.\,(\ref{Self-Energy}). Fig.\,\ref{Spectral Function} depicts an effective spectral function that takes into account the effect of a Fermi sea at half-filling in the adiabatic limit. It can be noticed that the bottom of the band reveals two parabolic band in this limit, in agreement with the two bands reported earlier in the single electron case.

Besides, the high-frequency results presented here are equivalent to the ones obtained in the framework of Floquet Green functions [\onlinecite{tsuji2008correlated}], whose definition relies on the time-dependent Hamiltonian in Eq.\,(\ref{Time-Dependent Hamiltonian}). Nevertheless, the Floquet Green functions are not based on the high-frequency assumption and enables us to numerically describe the effect of a slower driving frequency. The spectral function it leads to is illustrated in Fig.\,\ref{Spectral Function} for a frequency that satisfies $\Omega = \nu$. Out of equilibrium the energy is no longer a conserved quantity but, in the case of a time-periodic driving, Floquet's theory ensures that it is conserved up to a multiple of the frequency. This is the reason why the Floquet spectral function in Fig.\,\ref{Spectral Function} is similar to the effective one, but there are also replicas that are centered on $m\Omega$ for all values of the relative integer $m$. Actually, these replicas do exist in the high-frequency description too, but they can be neglected when the driving is off resonant.

\begin{figure}[t]
\centering
$\begin{array}{cc}
\includegraphics[trim = 00mm 0mm 0mm 0mm, clip, width=4.2cm]{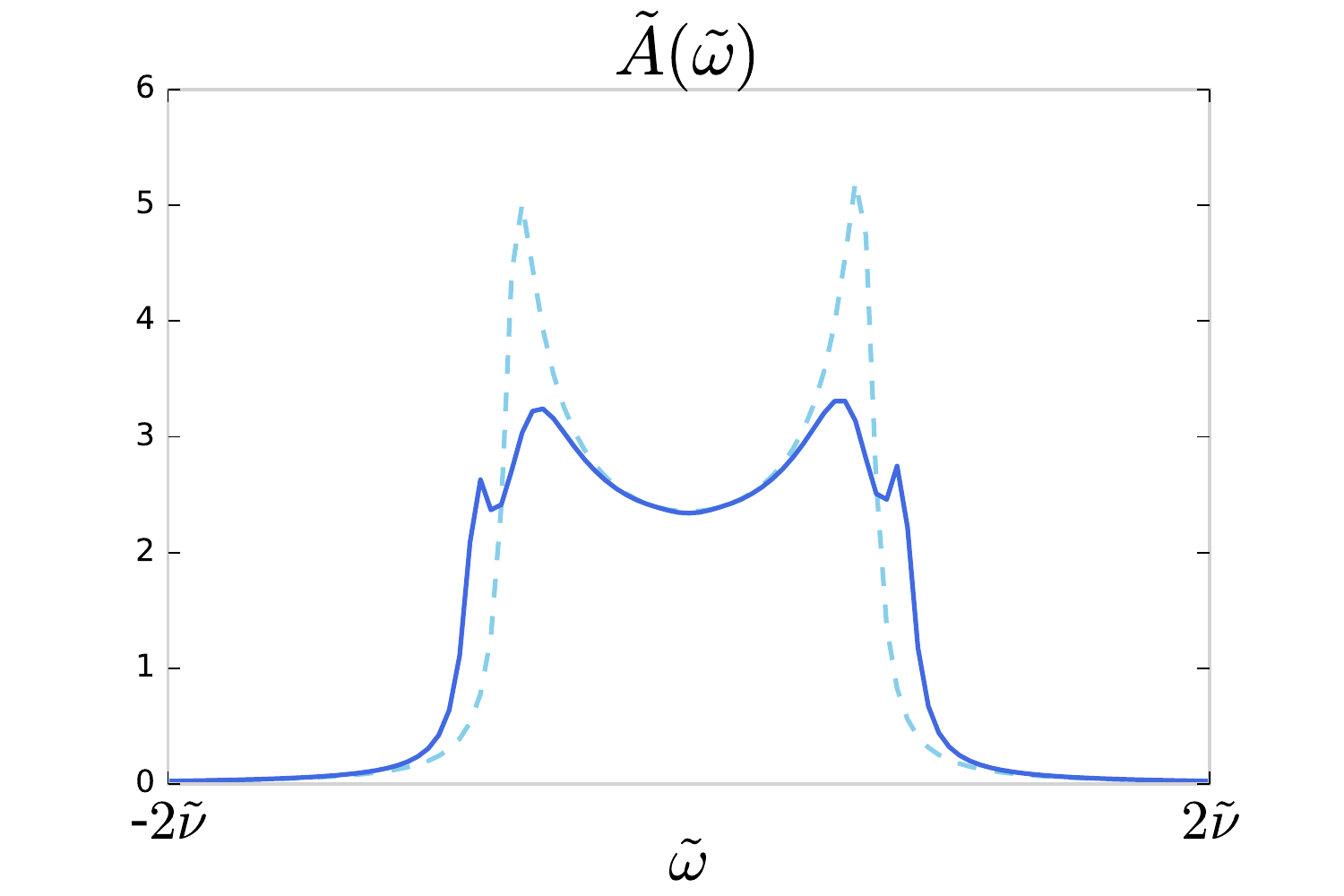} &
\includegraphics[trim = 00mm 0mm 0mm 0mm, clip, width=4.2cm]{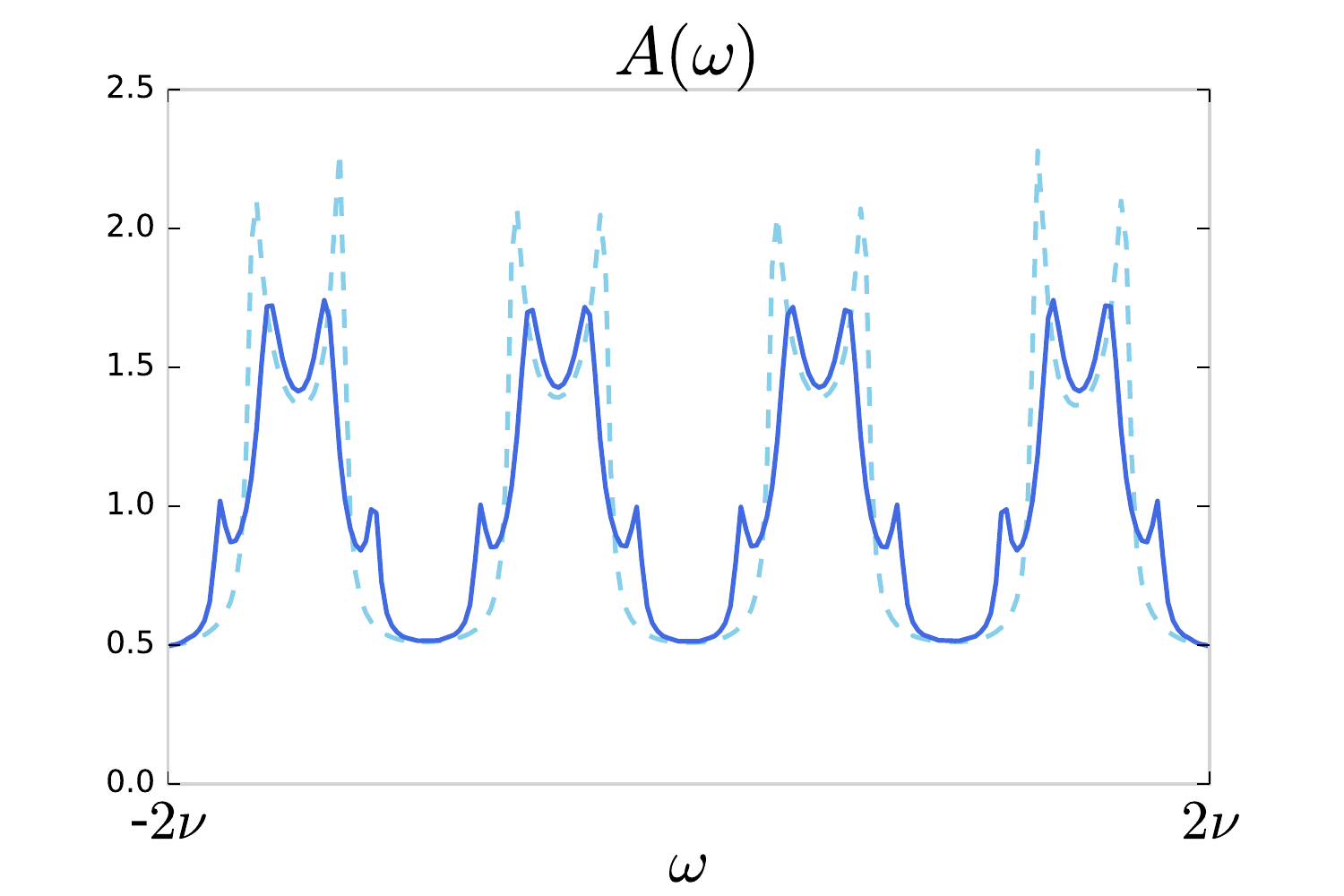}
\end{array}$
\caption{\small (Color online) Effective and Floquet local spectral functions for $\Omega = 5\nu$ (left) and $\Omega = \nu$ (right), respectively. Both plots corresponds to the case of zero temperature with $\omega_{0}=0.1\nu$, $z=1.8$, $\delta=0.01$ and $g_{0}=0.0\nu$ (dashed line) or $g_{0}=0.2\nu$ (full line).}
\label{DOS}
\end{figure}

The density of states, which is obviously a gauge invariant quantity too, can finally be obtained from the momentum integral of the spectral function over the Brillouin zone. It is depicted in Fig.\,\ref{DOS} in the adiabatic limit at zero temperature from both the high-frequency limit and Floquet Green functions. Whereas it shows a single band with polaronic peaks in the hight-frequency limite, there are additional replicas that overlap each other when reducing the driving frequency, in agreement with the Floquet spectral function in Fig.\,\ref{Spectral Function}.

%%%
\section{Strong-coupling regime}

\subsection{Lang-Firsov canonical transformation}
In equilibrium, the electron-phonon interaction may already be too large to be regarded as a perturbation with respect to the electron bandwidth. But regardless of the equilibrium interaction strength, we have also stressed that the system can always be dynamically driven toward such a strong-coupling regime defined by $|\tilde{t}_{1}(z)| \ll \tilde{g}_{0}$. This problem can be solved within a perturbation theory, whose zeroth order is given by $\tilde{t}_{1}(z)=0$ and usually describes localized electrons. This provides an exact analytical solution when the system lies in equilibrium, which is traditionally obtained from Lang-Firsov canonical transformation [\onlinecite{lang1962title}]. In our case, this transformation, which is detailed in Appendix, turns effective Hamiltonian (\ref{Effective Holstein Hamiltonians}) into
\begin{align}\label{Lang Firsov Hamiltonian}
\tilde{H}' &= \tilde{\omega}_{0} \sum_{q} \, b_{q}^{\dagger} b_{q}
- \tilde{\Delta} \sum_{n} c_{n}^{\dagger} c_{n} \\
&+ \tilde{t}_{1} \sum_{n} \left( c_{n+1}^{\dagger} c_{n} X_{n+1}^{\dagger}X_{n} + h.c. \right) \notag \\
&+ \tilde{t}_{2} \sum_{n} \left( c_{n+2}^{\dagger} c_{n} X_{n+2}^{\dagger}X_{n} + h.c. \right) \notag \\
&+ \tilde{g}_{2} \sum_{n,q} (2\cos q - 1) \, e^{-iqn} \left( c_{n+2}^{\dagger} c_{n} X_{n+2}^{\dagger}X_{n} + h.c. \right) B_{q} ~, \notag
\end{align}
where the polaron-polaron interactions are neglected and
\begin{align}
X_{n'}^{\dagger}X_{n} = \exp \left( \sum_{q} u_{q} \, (e^{-iqn}-e^{-iqn'})(b_{q}-b_{-q}^{\dagger}) \right)
\end{align}
with $u_{q}=[\tilde{g}_{0}+(2\cos q - 1)\tilde{g}_{1}]/\tilde{\omega}_{0}$.

Whereas the phonon frequency is not changed by the canonical transformation, the polaron binding energy
\begin{align}\label{Binding Energy Strong Coupling}
\tilde{\Delta}(z) = \frac{\tilde{g}_{0}^{2}-4\tilde{g}_{0}\tilde{g}_{1}(z)}{\tilde{\omega}_{0}}
\end{align}
is reduced by Peierls coupling $\tilde{g}_{1}$ when the driving is turned on, which is illustrated in Fig.\,\ref{Renormalized Binding Energy}. It defines a potential well that aims to localize the electron onto a molecular site, so that the characteristic size of the polaron becomes comparable to the lattice scale, hence the name of small polaron that may be encountered sometimes in the litterature. Note that $\tilde{\Delta}$ does not change signs because $\tilde{g}_{1}$ comes as a second-order correction to $\tilde{g}_{0}$ in the high-frequency limit, accordingly to Eq.\,(\ref{Coupling Definitions}).
Of course, one naturally recovers the equilibrium binding energy when the driving is turned off ($z=0$). In this case, the binding energies introduced in the strong-coupling regime and in the non-adiabatic limit of the weak-coupling regime equal each other [\onlinecite{klamt1988tight},\,\onlinecite{barivsic2008phase}]. Interestingly, this is no longer the case out of equilibrium, as it can be seen from Eq.\,(\ref{Binding Energy NonAdiabatic}) and Eq.\,(\ref{Binding Energy Strong Coupling}). The extra term $4\tilde{g}_{0}\tilde{g}_{2}(z)$ in Eq.\,(\ref{Binding Energy NonAdiabatic}) comes from the phonon-assisted next-nearest hopping process which leads to $4\tilde{g}_{0}\tilde{g}_{2}(z)\cos(2k)$ in momentum space (cf. non-adiabatic limit in Appendix) and whose expansion for small $k$ yields an energy off-set.

Contrary to the equilibrium situation, the canonical transformation does not diagonalize the effective Hamiltonian when the off-resonant driving turns off the nearest-neighbor hopping, i.e. when $t_{1}(z) = 0$. This is due to nonequilibrium coupling $\tilde{g}_{2}$ that is responsible for the two last terms in the right-hand side of Eq.\,(\ref{Lang Firsov Hamiltonian}). The first one, which scales with
\begin{align}
\tilde{t}_{2}(z) = 2\frac{\tilde{g}_{0}\tilde{g}_{2}(z)}{\tilde{\omega_{0}}}~,
\end{align}
describes the next-nearest-neighbor hopping of the polaron, namely the electron dressed by the phonon cloud whose annihilation operator is $c_{n} X_{n}$. This hopping process tends to delocalize the polaron and competes the nearest-neighbor hopping when $\tilde{t}_{1}\sim \tilde{t}_{2}$, which roughly occurs when
\begin{align}
\frac{\nu}{\omega_{0}} \sim \left( \frac{\Omega}{g_{0}} \right)^{2} ~.
\end{align}
Such a condition is for example accessible in the adiabatic situation where $\omega_{0}\ll \nu$.
The second term generated by nonequilibrium coupling $\tilde{g}_{2}$ in Eq.\,(\ref{Lang Firsov Hamiltonian}) describes phonon-assisted polaron hopping between next-nearest-neighbor sites.

\subsection{Peierls-Feynman-Bogoliubov variational principle}

In order to get rid of the phonon-assisted polaron hopping term in Hamiltonian (\ref{Lang Firsov Hamiltonian}), we aim to map it onto
\begin{align}\label{Quadratic Hamiltonian}
H^{*} &= \tilde{\omega}_{0} \sum_{q} b_{q}^{\dagger}b_{q} 
- \tilde{\Delta} \sum_{n}c^{\dagger}_{n}c_{n} \notag \\
&+ t_{1}^{*} \sum_{n} \left(c^{\dagger}_{n+1}c_{n}+h.c.\right)  + t_{2}^{*} \sum_{n} \left(c^{\dagger}_{n+2}c_{n}+h.c.\right) ~.
\end{align}
This Hamiltonian is quadratic in momentum space, so that we know its partition function $Z^{*}=\Tr e^{-\beta H^{*}}$. Parameters $t_{1}^{*}$ and $t_{2}^{*}$ are then determined under the constraint that $\rho^{*} = \Tr e^{-\beta H^{*}}/Z^{*}$ is the best approximation of the exact density operator defined from Hamiltonian $\tilde{H}'$. This leads to Peierls-Feynman-Bogoliubov variational principle [\onlinecite{PhysRev.54.918,Bogolyubov,feynman1972lectures}], which consists in minimizing, with respect to $t_{1}^{*}$ and $t_{2}^{*}$, the following functional
\begin{align}
F^{*}+\langle \tilde{H}' - H^{*} \rangle_{*} ~,
\end{align}
where $F^{*} = - (1/\beta) \ln Z^{*}$. This results in
\begin{align}
t_{1}^{*} = \tilde{t}_{1} \left\langle X_{m+1}^{\dagger}X_{m} \right\rangle_{*}
~~~~~~\text{and}~~~~~~
t_{2}^{*} = \tilde{t}_{2} \left\langle X_{m+2}^{\dagger}X_{m} \right\rangle_{*} ~,
\end{align}
The reader may find more details about the derivation of these expressions in Appendix.

\subsection{Holstein polaron band}

\begin{figure}[t]
\centering
$\begin{array}{cc}
\includegraphics[trim = 8mm 0mm 10mm 5mm, clip, width=4.2cm]{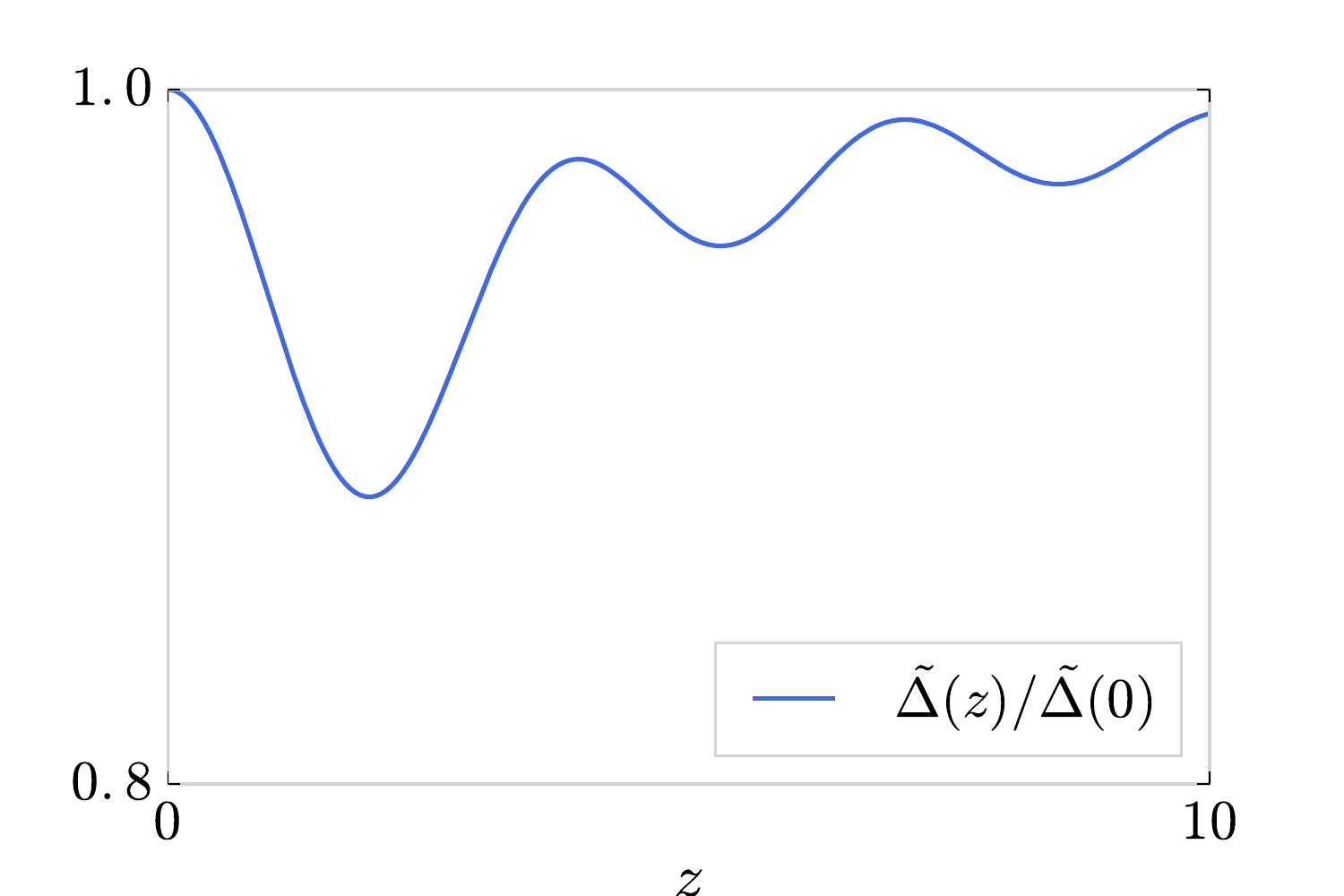} &
\includegraphics[trim = 8mm 0mm 10mm 5mm, clip, width=4.2cm]{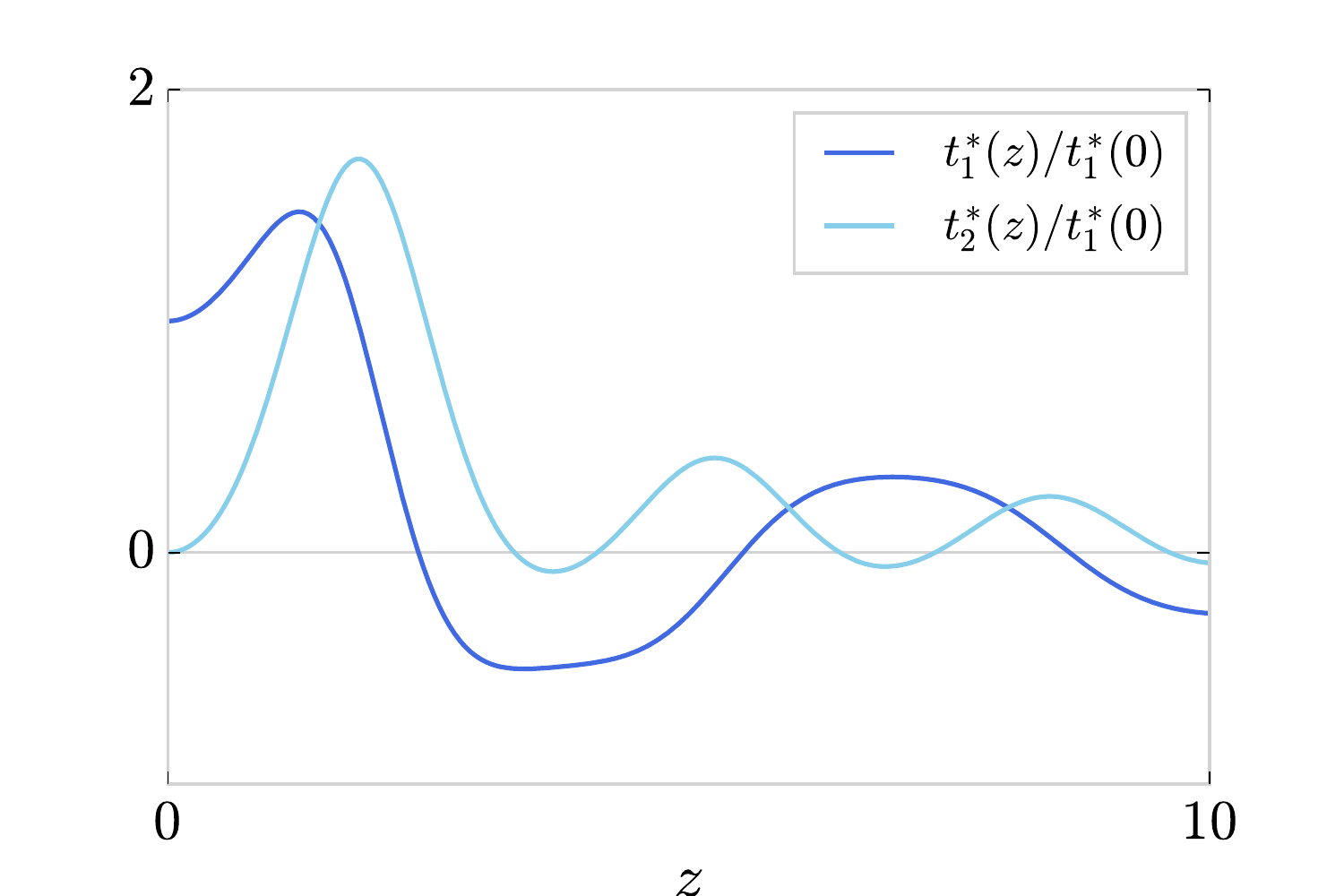}
\end{array}$
\caption{\small (Color online) Variations of the polaron binding energy (left) and of its nearest- and next-nearest-neighbor hopping amplitudes (right) for $\Omega = 5\nu$, $g_{0}=\nu$, $\omega_{0}=0.1\,\nu$, and zero temperature.}
\label{Renormalized Binding Energy}
\end{figure}

It is worth mentioning that the variational principle simply relies on the averages of bosonic operators, meaning that it describes hopping processes that conserve the number of phonons. If this elastic process is dominant, then the electron remains coherent and can still be described in terms of Bloch band theory. The average of bosonic operators can be evaluated from Feynman disentangling method, which is detailed in Appendix. The result is
\begin{align}
\left\langle X_{m+n}^{\dagger}X_{m} \right\rangle_{*}
&= \exp \left( - (2N_{0}+1)\frac{\tilde{g}_{0}^{2}-4\tilde{g}_{0}\tilde{g}_{1}-2\tilde{g}_{0}\tilde{g}_{1}\delta_{n,1}}{\tilde{\omega}_{0}^{2}} \right)
\end{align}
so that the nearest- and next-nearest-neighbor hopping amplitudes are functions of the phonon temperature and the driving strength. They are respectively given by
\begin{align}\label{Polaron Assisted Hopping 1}
t_{1}^{*}(z) &= \tilde{t}_{1}(z) \exp \left( - (2N_{0}+1)\frac{\tilde{g}_{0}^{2}-6\tilde{g}_{0}\tilde{g}_{1}(z)}{\tilde{\omega}_{0}^{2}} \right) ~~ 
\end{align}
and
\begin{align}\label{Polaron Assisted Hopping 2}
t_{2}^{*}(z) &= \tilde{t}_{2}(z) \exp \left( - (2N_{0}+1)\frac{\tilde{g}_{0}^{2}-4\tilde{g}_{0}\tilde{g}_{1}(z)}{\tilde{\omega}_{0}} \right) ~.
\end{align}

These hopping processes both tend to delocalize the electron and thus compete the potential well $\tilde{\Delta}$ to enhance the polaron size. The largest polaron it characterizes is expected to be found at low temperatures where phonon occupation number $N_{0}$ vanishes. When increasing the temperature, the electron bandwidth becomes flatter and flatter so that its effective mass becomes heavier and heavier. Therefore, the inelastic processes, which do not conserve the number of phonons, become more and more important. The electron loses its coherence and get a diffusive motion. However, these effects are only due to the existence of polarons in the sense that they already occur in equilibrium without time-periodic driving.

The nonequilibirum effects due to the off-resonant driving are actually double. On the one hand, it yields next-nearest-neighbor hopping processes which cannot be switched off dynamically together with the nearest-neighbor ones, i.e. the conditions $\tilde{t}_{1}(z)=0$ and $\tilde{t}_{2}(z)=0$ cannot be satisfied simultaneously. This is what Fig.\,\ref{Renormalized Binding Energy} illustrates. As a consequence, the dynamical localization of electrons predicted in Ref. [\onlinecite{PhysRevB.34.3625}] no longer arises in the presence of lattice vibrations. On the other hand, nonequilibrium Peierls coupling $\tilde{g}_{1}$ enhances the exponential arguments in Eqs.\,(\ref{Polaron Assisted Hopping 1}) and (\ref{Polaron Assisted Hopping 2}). This is reason why $t_{1}(z)$ first becomes larger when turning on the driving strength in Fig.\,\ref{Renormalized Binding Energy}. The driving-renormalized polaron band it leads to finally satisfies
\begin{align}
\xi_{k}^{*}(z) = 2t_{1}^{*}(z)\cos(k) + 2 t_{2}^{*}(z)\cos(2k) - \tilde{\Delta}(z) ~.
\end{align}
Thus, contrary to the equilibrium case, the polaron is also allowed to dynamically enhance the electronic bandwidth and reduce the effective mass of the electron.

%%%
\section{Conclusion}

Here we have addressed the problem of rapidly driven electrons that are linearly coupled to vibrational modes in a one-dimensional crystal. The stroboscopic dynamics has been apprehended up to the third-order expansion in the high-frequency limit. This approach provides an effective description of the problem in term of a time-independent effective Hamiltonian. It has enabled us to show that any kind of electron-phonon interaction is responsible for corrections to the effective Hamiltonian which reduces the interaction strength between electrons and phonons of specific momenta. In this sense, the off-resonant driving can be regarded as a way to tune the electron-phonon coupling and to chose specific interaction channels in a dynamical and reversible fashion.

Finally, we have discussed the specific case of Holstein interaction in equilibrium. Such a local interaction is responsible for non-local interactions when the electrons are rapidly driven, such as antisymmetric interactions of Peierls type and phonon-assisted electron tunneling, which suppresses the dynamical Wannier-Stark localization. The polaronic effects these nonequilibrium corrections induce have been reported in the weak- and strong-coupling regimes, since these two regimes can both be visited dynamically when varying the driving strength. In particular, we have described how the binding energy, the mass and the size of the polaron may be controlled by the off-resonant driving. These high-frequency results have also been compared to the ones obtained in the formalism of Floquet Green functions, which allows the description of driving with arbitrary (low) frequencies.

Although the high-frequency limit is already relevant for systems such as shaken optical lattices, the explicit knowledge of the electron-phonon mechanisms we derive here in the third-order expansion allows the description of slower frequencies that become reasonable for solid state physics too, for example during multicycle laser irradiations in pump-probe experiments. The dynamical control allowed by the driving strength offers the possibility to test weak- and strong-coupling polaron theories within a single material and may also be helpful to understand the crucial interplay between local and nonlocal electron-phonon interactions in systems such as organic molecular crystals.

\begin{acknowledgments}
The authors are grateful to E. A. Stepanov and would like to point out his involvement into the derivation of the effective Green function formalism. This work was supported by NWO via Spinoza Prize and by ERC Advanced Grant 338957 FEMTO/NANO.
\end{acknowledgments}

\bibliographystyle{apsrev4-1}
\bibliography{references}

\newpage
\onecolumngrid

\appendix

%%%
\section{Third-order correction to the effective Hamiltonian}\label{Appendix Third-order correction to the effective Hamiltonian}

The third-order correction in the high-frequency limit is given by
\begin{align}
\tilde H_{3} &= \frac{1}{2}\sum_{m\neq0} \sum_{k,k'} \frac{\epsilon_{k,m} \epsilon_{k',-m}}{m^{2}} [ [ c^{\dagger}_{k}c_{k}, H_{ep} ], c^{\dagger}_{k'}c_{k'} ] ~.
\end{align}
It relies on the following commutations:
\begin{align}
\left[ c^{\dagger}_{k}c_{k},  c^{\dagger}_{k''+q}c_{k''} \right] &= c^{\dagger}_{k''+q}c_{k''} \left( \delta_{k,k''+q} - \delta_{k,k''} \right)
\end{align}
and
\begin{align}
\left[ \left[ c^{\dagger}_{k}c_{k}, c^{\dagger}_{k''+q}c_{k''} \right], c^{\dagger}_{k'}c_{k'} \right]
&= \left[c^{\dagger}_{k''+q}c_{k''}, c^{\dagger}_{k'}c_{k'} \right] \left( \delta_{k,k''+q} - \delta_{k,k''} \right) \notag \\
&= c^{\dagger}_{k''+q}c_{k''} \left( \delta_{k,k''+q} - \delta_{k,k''} \right) \left( \delta_{k',k''} - \delta_{k',k''+q} \right) \notag \\
&= c^{\dagger}_{k''+q}c_{k''} \left( \delta_{k,k''+q}\delta_{k',k''} + \delta_{k,k''}\delta_{k',k''+q} -\delta_{k,k''+q}\delta_{k',k''+q} - \delta_{k,k''}\delta_{k',k''} \right) ~,
\end{align}
which subsequently leads to
\begin{align}
\tilde{H}_{3} = - \sum_{k,q} \frac{g_{q}}{\delta E} ~ \eta_{k,q}(z) ~ c^{\dagger}_{k+q}c_{k}B_{q} ~.
\end{align}
Thus the third-order correction yields an additional electron-phonon coupling whose momentum dependence is characterized by
\begin{align}
\eta_{k,q} &= \frac{1}{2} \sum_{m \neq 0} \frac{1}{m^{2}} \left( \epsilon_{k+q,m}\epsilon_{k+q,-m} + \epsilon_{k,m}\epsilon_{k,-m} - \epsilon_{k+q,m}\epsilon_{k,-m} - \epsilon_{k,m}\epsilon_{k+q,-m} \right) \notag \\
&= \sum_{m > 0} \frac{1}{m^{2}} \left( \epsilon_{k+q,m}\epsilon_{k+q,-m} + \epsilon_{k,m}\epsilon_{k,-m} - \epsilon_{k+q,m}\epsilon_{k,-m} - \epsilon_{k,m}\epsilon_{k+q,-m} \right) ~.
\end{align}
Because it involves products of two opposite harmonics of the electronic dispersion relation, the electron-phonon coupling becomes $k$-dependent. Besides, these harmonics are defined as
\begin{align}
\epsilon_{k,m}(z) &= \int_{-\pi}^{+\pi} \frac{dt}{2\pi} e^{imt}\epsilon_{k}(\tau) = -\frac{1}{2} J_{m}(z) \left( e^{ik}+\left(-1\right)^{m}e^{-ik} \right) = J_{m}(z)
\left |
\begin{aligned}
\epsilon_{k} \\
i\bar{\epsilon}_{k}
\end{aligned}
\right . ~,
\end{align}
where $\epsilon_{k}=\frac{2\nu}{\delta E}\cos(k)$ and $\bar{\epsilon}_{k}=\frac{2\nu}{\delta E}\sin(k)$ refer to even and odd values of $m$, respectively. As a result
\begin{align}
\epsilon_{k,m}(z) \epsilon_{k+q,-m}(z) &= \epsilon_{k+q,m}(z) \epsilon_{k,-m}(z) 
= J_{m}^{2}(z)
\left |
\begin{aligned}
\epsilon_{k} \epsilon_{k+q} \\
\bar{\epsilon}_{k} \bar{\epsilon}_{k+q}
\end{aligned}
\right .
\end{align}
and finally
\begin{align}
\eta_{k,q}(z) = \sum_{m>0} \frac{J^{2}_{2m}(z)}{(2m)^{2}} \left( \epsilon_{k+q} - \epsilon_{k} \right)^{2} + \sum_{m>0} \frac{J^{2}_{2m-1}(z)}{(2m-1)^{2}} \left( \bar{\epsilon}_{k+q} - \bar{\epsilon}_{k} \right)^{2} ~.
\end{align}
Since this function is strictly positive for non-vanishing fields, the off-resonant driving of electrons essentially reduces the electron-phonon interaction for some specific values of $k$ and $q$, so that it can be used to dynamically couple specific electrons and phonons.

%%%
\section{Electron-phonon interactions in real space}\label{Appendix Electron-phonon interactions in real space}

Within the third-order description of the high-frequency limit, the renormalized electron-phonon coupling is
\begin{align}
\gamma_{k,q}(z) &= \frac{g_{q}}{\delta E} \left( 1 - \eta_{k,q} \left( z \right) \lambda^{2} \right) ~.
\end{align}
In order to highlight what kinds of electron-phonon interactions the off-resonant driving generates, it is quite instructive to rephrase this coupling in terms of real space coordinates. To do so, it is convenient to first linearize the following terms
\begin{align}
\left( \epsilon_{k} - \epsilon_{k+q} \right)^{2} 
&= \left( \epsilon_{k} - \epsilon_{q}\epsilon_{k} + \bar{\epsilon}_{q}\bar{\epsilon}_{k} \right)^{2}  \notag \\
&= \left( 1 - \epsilon_{q} \right)^{2} \epsilon_{k}^{2} + 2 \left( 1 - \epsilon_{q} \right) \bar{\epsilon}_{q} \epsilon_{k} \bar{\epsilon}_{k} + \bar{\epsilon}_{q}^{2} \bar{\epsilon}_{k}^{2} \notag \\
&= \left( 1 - 2\epsilon_{q} + \epsilon_{q}^{2} \right) \left( \frac{1}{2}+\frac{1}{2}\epsilon_{2k}  \right) + \left( 1 - \epsilon_{q} \right)\bar{\epsilon}_{q}\bar{\epsilon}_{2k} +\bar{\epsilon}_{q}^{2} \left( \frac{1}{2}-\frac{1}{2}\epsilon_{2k}  \right) \notag \\
&= \left( 1-\epsilon_{q} \right) + \frac{1}{2} \left( 1 - 2\epsilon_{q} + \epsilon_{q}^{2} - \bar{\epsilon}_{q}^{2} \right) \epsilon_{2k} + \left( \bar{\epsilon}_{q} - \frac{1}{2}\bar{\epsilon}_{2q} \right)\bar{\epsilon}_{2k} \notag \\
&= \left( 1-\epsilon_{q} \right) + \frac{1}{2} \left( 1 - 2\epsilon_{q} + \epsilon_{2q} \right) \epsilon_{2k} + \left( \bar{\epsilon}_{q} - \frac{1}{2}\bar{\epsilon}_{2q} \right)\bar{\epsilon}_{2k}
\end{align}
and
\begin{align}
\left( \bar{\epsilon}_{k} - \bar{\epsilon}_{k+q} \right)^{2} 
&= \left( \epsilon_{k-\pi/2} - \epsilon_{k+q-\pi/2} \right)^{2}  \\
&= \left( 1-\epsilon_{q} \right) + \frac{1}{2} \left( 1 - 2\epsilon_{q} + \epsilon_{2q} \right) \epsilon_{2k-\pi} + \left( \bar{\epsilon}_{q} - \frac{1}{2}\bar{\epsilon}_{2q} \right)\bar{\epsilon}_{2k-\pi} \notag \\
&= \left( 1-\epsilon_{q} \right) - \frac{1}{2} \left( 1 - 2\epsilon_{q} + \epsilon_{2q} \right) \epsilon_{2k} - \left( \bar{\epsilon}_{q} - \frac{1}{2}\bar{\epsilon}_{2q} \right)\bar{\epsilon}_{2k} ~. \notag
\end{align}
It is now assumed that $\epsilon_{x}=\cos x$ and $\bar{\epsilon}_{x}=\sin x$ for more convenience. Then, the renormalized electron-phonon coupling can be rewritten as
\begin{align}
\gamma_{k,q}(z) &= \frac{g_{q}}{\delta E} \left[ 1 -  \sigma(z) ~ \lambda^{2} \left(\frac{2\nu}{\delta E}\right)^{2} ~ \left( 1-\epsilon_{q} \right) + \frac{\delta(z)}{2} ~ \lambda^{2}\left(\frac{2\nu}{\delta E}\right)^{2} ~ \left[ \left( 1 - 2\epsilon_{q} + \epsilon_{2q} \right) ~ \epsilon_{2k} + \left( 2\bar{\epsilon}_{q} - \bar{\epsilon}_{2q} \right)\bar{\epsilon}_{2k} \right] \right]
\end{align}
with
\begin{align}
\sigma(z) = \sum_{m>0}\frac{J_{m}^{2}(z)}{m^{2}} ~~~ \text{and} ~~~ \delta(z) = \sum_{m>0} \left( \frac{J_{2m-1}^{2}(z)}{(2m-1)^{2}} - \frac{J_{2m}^{2}(z)}{(2m)^{2}} \right) ~.
\end{align}
Regardless of the energy scale involved in the definition of the small parameter $\lambda$, only $\lambda^{2}(2\nu)^{2}/\delta E^{2}=(2\nu/\Omega)^{2}$ is relevant for the renormalized electron-phonon interaction. This is understandable because the third-order corrections only arise from harmonics of the electronic dispersion relation, whose characteristic energy scale is $2\nu$. 
The effective electron-phonon Hamiltonian is then defined as
\begin{align}
\tilde{H}_{e-p} &= \lambda \sum_{k,q} \gamma_{k,q}  ~ c^{\dagger}_{k+q}c_{k}B_{q} \notag \\
&= \lambda \sum_{l,m,n} \sum_{k,q} \gamma_{l,q} ~ c^{\dagger}_{m}c_{n}B_{q} ~ e^{ik(l-m+n)}e^{-iqm} \notag \\
&= \lambda \sum_{m,n} \sum_{q} \gamma_{m-n,q} ~ c^{\dagger}_{m}c_{n}B_{q} ~ e^{-iqm} \notag\\
&= \lambda \sum_{m,n} \sum_{\mu,\nu} \sum_{q} \gamma_{m-n,\nu} ~ c^{\dagger}_{m}c_{n}B_{\mu} ~ e^{iq(\mu-m+\nu)} \notag \\
&= \lambda \sum_{l,m,n} \gamma_{m-n,m-l} ~ c^{\dagger}_{m}c_{n}B_{l} ~.
\end{align}
For a local electron-phonon coupling $g_{0}$ in equilibrium, the renormalized coupling out of equilibrium satisfies
\begin{align}
\gamma_{m-n,m-l} (z) &= \frac{g_{0}}{\delta E} ~ \delta_{m,n} \, \delta_{l,m}  \notag \\
&+ \frac{g_{0}}{\delta E} \left( \frac{2\nu}{\Omega} \right)^{2} \frac{\sigma(z)}{2} ~ \delta_{m,n}\left(\delta_{l,m-1}-2\delta_{l,m}+\delta_{l,m+1}\right) \notag \\
&+ \frac{g_{0}}{\delta E} \left( \frac{2\nu}{\Omega} \right)^{2} \frac{\delta(z)}{4} ~ \left[ \delta_{m-2,n}\left(\delta_{l,m}-2\delta_{l,m-1}+\delta_{l,m-2} \right) + \delta_{m+2,n}\left(\delta_{l,m}-2\delta_{l,m+1}+\delta_{l,m+2} \right) \right] ~.
\end{align}
The effective electron-phonon Hamiltonian in real space is finally rewritten as:
\begin{align}
\tilde{H}_{e-ph}
&= \tilde{g}_{0} \sum_{m} c^{\dagger}_{m}c_{m}B_{m} \notag \\
&+ \tilde{g}_{1}(z) \sum_{m}c^{\dagger}_{m}c_{m}\left(B_{m+1}-2B_{m}+B_{m-1}\right) \notag \\
&+ \tilde{g}_{2}(z) \sum_{m}c^{\dagger}_{m+2}c_{m}\left(B_{m+2}-2B_{m+1}+B_{m}\right) + h.c. ~,
\end{align}
where the dimensionless electron-phonon couplings are defined as
\begin{align}
\tilde{g}_{0} &= \frac{g_{0}}{\Omega} \notag \\
\tilde{g}_{1}(z) &= \frac{g_{0}}{\Omega} \left( \frac{2\nu}{\Omega} \right)^{2} \frac{\sigma(z)}{2} \notag \\
\tilde{g}_{2}(z) &= \frac{g_{0}}{\Omega} \left( \frac{2\nu}{\Omega} \right)^{2} \frac{\delta(z)}{4} ~.
\end{align}

%%%
\section{Retarded component of the effective self-energy}\label{Appendix Retarded component of the effective self-energy}

\begin{figure}[t]
\centering
$\begin{array}{c}
\includegraphics[trim = 0mm 0mm 0mm 0mm, clip, width=8cm]{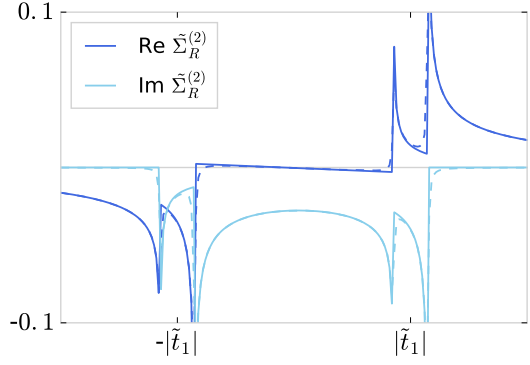}
\end{array}$
\caption{\small (Color online) Real and imaginary parts of the retarded component of the effective self-energy for a single electron at room temperature. Analytics (full lines) is compared to numerics (dashed lines) for $\Omega = 5\nu$, $\omega_{0}=0.1\nu$, $g_{0}=0.2\nu$, $z=1.8$, $\delta=0.01$ and $k=0$.}
\label{Appendix SelfEnergy}
\end{figure}

From the second-order perturbation theory introduced in the main text, the retarded component of the effective self-energy is
\begin{align}
\tilde{\Sigma}^{(2)}_{R}(k,\tilde{\omega}) &= \int_{BZ}dq \, \gamma_{k,q}\gamma_{k+q,-q} \left( \frac{N_{0}+n_{k+q}}{\tilde{\omega} - 2\tilde{t}_{1}\epsilon_{k+q} + \tilde{\omega}_{0} + i0^{+}} +\frac{N_{0}+1-n_{k+q}}{\tilde{\omega} - 2\tilde{t}_{1}\epsilon_{k+q} - \tilde{\omega}_{0} + i0^{+}} \right)  ~,
\end{align}
where $N_{0}$ is the equilibrium distribution function of dispersionless phonons, $n_{k}=1/(1+e^{\beta2\tilde{t}_{1}\epsilon_{k}})$ and 
\begin{align}
\gamma_{k,q}\gamma_{k+q,-q} &= |\gamma_{k,q}|^{2} \notag \\
&=\tilde{g}_{0}^{2} \left( 1 - 2\eta_{k,q} \lambda^{2} \right) + o(\lambda^{3})
\end{align}
The imaginary part of the self-energy relies on the following integral:
\begin{align}
I &= - \Imag \int_{BZ}dq \frac{|\gamma_{k,q}|^{2}n_{k+q}}{\tilde{\omega} \pm \tilde{\omega}_{0} + i0^{+} - 2\tilde{t}_{1} \epsilon_{k+q}} \notag \\
&= - \tilde{g}_{0}^{2} \Imag \int_{BZ}dq \frac{n_{k+q}}{\tilde{\omega} \pm \tilde{\omega}_{0} + i0^{+} - 2\tilde{t}_{1} \epsilon_{k+q}}
- 4\tilde{g}_{0}\tilde{g}_{1} \Imag \int_{BZ}dq \frac{n_{k+q}(\epsilon_{q}-1)}{\tilde{\omega} \pm \tilde{\omega}_{0} + i0^{+} - 2\tilde{t}_{1} \epsilon_{k+q}} \notag \\
&- 4\tilde{g}_{0}\tilde{g}_{2} \, \epsilon_{2k} \Imag \int_{BZ}dq \frac{ n_{k+q}( 1 - 2\epsilon_{q} + \epsilon_{2q})}{\tilde{\omega} \pm \tilde{\omega}_{0} + i0^{+} - 2\tilde{t}_{1} \epsilon_{k+q}}
- 4\tilde{g}_{0}\tilde{g}_{2} \, \bar{\epsilon}_{2k} \Imag \int_{BZ}dq \frac{ n_{k+q}( 2\bar{\epsilon}_{q} - \bar{\epsilon}_{2q}) }{\tilde{\omega} \pm \tilde{\omega}_{0} + i0^{+} - 2\tilde{t}_{1} \epsilon_{k+q}} \notag \\
&= \tilde{g}_{0}^{2} \int_{-1}^{1} \frac{d\epsilon_{q}}{\sqrt{1-\epsilon_{q}^{2}}} ~ n_{q} \, \delta\left(\tilde{\omega} \pm \tilde{\omega}_{0} - 2\tilde{t}_{1} \epsilon_{q}\right)
+ 4\tilde{g}_{0}\tilde{g}_{1} \int_{-1}^{1} \frac{d\epsilon_{q}}{\sqrt{1-\epsilon_{q}^{2}}} ~ n_{q} \, (\epsilon_{k}\epsilon_{q}-1)  \, \delta\left(\tilde{\omega} \pm \tilde{\omega}_{0} - 2\tilde{t}_{1} \epsilon_{q}\right) \notag \\
&+ 4\tilde{g}_{0}\tilde{g}_{2} \, \epsilon_{2k} \int_{-1}^{1} \frac{d\epsilon_{q}}{\sqrt{1-\epsilon_{q}^{2}}} ~ n_{q} \, \left(1 - 2\epsilon_{k}\epsilon_{q} + \epsilon_{2k}(2\epsilon_{q}^{2}-1) \right) \, \delta\left(\tilde{\omega} \pm \tilde{\omega}_{0} - 2\tilde{t}_{1} \epsilon_{q}\right) \notag \\
&+ 4\tilde{g}_{0}\tilde{g}_{2} \, \bar{\epsilon}_{2k} \int_{-1}^{1} \frac{d\epsilon_{q}}{\sqrt{1-\epsilon_{q}^{2}}} ~ n_{q} \, \left( - 2\bar{\epsilon}_{k}\epsilon_{q} + \bar{\epsilon}_{2k}(2\epsilon_{q}^{2}-1) \right) \, \delta\left(\tilde{\omega} \pm \tilde{\omega}_{0} - 2\tilde{t}_{1} \epsilon_{q}\right) \notag \\
&=  n_{X_{\pm}} \left(
\tilde{g}_{0}^{2} - 4\tilde{g}_{0}\tilde{g}_{1} - 4\tilde{g}_{0}\tilde{g}_{2} + 4\tilde{g}_{0}\tilde{g}_{2}\epsilon_{2k} + 4\tilde{g}_{0}(\tilde{g}_{1}-2\tilde{g}_{2})\,\epsilon_{k}\,X_{\pm} + 8\tilde{g}_{0}\tilde{g}_{2}\,X_{\pm}^{2}
\right) \, \frac{\Theta\left(1-|X_{\pm}|\right)}{\sqrt{1-X_{\pm}^{2}}} ~,
\end{align}
where $n_{X}=1/(1+e^{\beta2\tilde{t}_{1}X})$, $X_{\pm}=(\tilde{\omega} \pm \tilde{\omega}_{0})/2\tilde{t}_{1}$ and $\Theta$ denotes the Heaviside step function. This leads to the imaginary part of the self-energy, namely
\begin{align}
\Imag \tilde{\Sigma}^{(2)}_{R}(k,\tilde{\omega})
=
&-\left[
N_{0}+n_{X_{+}}
\right]
\left[
\tilde{g}_{0}^{2} - 4\tilde{g}_{0}\tilde{g}_{1} - 4\tilde{g}_{0}\tilde{g}_{2} + 4\tilde{g}_{0}\tilde{g}_{2}\epsilon_{2k}
+ 4\tilde{g}_{0}(\tilde{g}_{1}-2\tilde{g}_{2})\,\epsilon_{k}\,X_{+} 
+ 8\tilde{g}_{0}\tilde{g}_{2}\,X_{+}^{2}
\right] \, 
\frac{\Theta\left(1-|X_{+}|\right)}{\sqrt{1-X_{+}^{2}}} \notag \\
&-
\left[
N_{0}+1-n_{X_{-}}
\right]
\left[
\tilde{g}_{0}^{2} - 4\tilde{g}_{0}\tilde{g}_{1} - 4\tilde{g}_{0}\tilde{g}_{2} + 4\tilde{g}_{0}\tilde{g}_{2}\epsilon_{2k} 
+ 4\tilde{g}_{0}(\tilde{g}_{1}-2\tilde{g}_{2})\,\epsilon_{k}\,X_{-} 
+ 8\tilde{g}_{0}\tilde{g}_{2}\,X_{-}^{2}
\right] \, 
\frac{\Theta\left(1-|X_{-}|\right)}{\sqrt{1-X_{-}^{2}}} ~.
\end{align}
The real part of the self-energy can then be obtained from Kramers-Kronig relation
\begin{align}
\Real \tilde{\Sigma}^{(2)}_{R}(k,\tilde{\omega})
&=
\int_{-\infty}^{+\infty} \frac{dX_{+}'}{X_{+}'-X_{+}}
\left[
N_{0}+n_{X_{+}'}
\right]
\left[
\tilde{g}_{0}^{2} - 4\tilde{g}_{0}\tilde{g}_{1} - 4\tilde{g}_{0}\tilde{g}_{2} + 4\tilde{g}_{0}\tilde{g}_{2}\epsilon_{2k} 
+ 4\tilde{g}_{0}(\tilde{g}_{1}-2\tilde{g}_{2})\,\epsilon_{k}\,X_{+}' 
+ 8\tilde{g}_{0}\tilde{g}_{2}\,X_{+}'^{2}
\right] \, 
\frac{\Theta\left(1-|X_{+}'| \right)}{\sqrt{1-X_{+}'^{2}}} \notag \\
&+
\int_{-\infty}^{+\infty} \frac{dX_{-}'}{X_{-}'-X_{-}}
\left[
N_{0}+1-n_{X_{-}'}
\right]
\left[
\tilde{g}_{0}^{2} - 4\tilde{g}_{0}\tilde{g}_{1} - 4\tilde{g}_{0}\tilde{g}_{2} + 4\tilde{g}_{0}\tilde{g}_{2}\epsilon_{2k} 
+ 4\tilde{g}_{0}(\tilde{g}_{1}-2\tilde{g}_{2})\,\epsilon_{k}\,X_{-}'
+ 8\tilde{g}_{0}\tilde{g}_{2}\,X_{-}'^{2}
\right] \, 
\frac{\Theta\left(1-|X_{-}'|\right)}{\sqrt{1-X_{-}'^{2}}} ~. \notag 
\end{align}
For a single electron in the band, it reduces to
\begin{align}\label{Appendix Real Self Energy Single Electron}
\Real \tilde{\Sigma}^{(2)}_{R}(k,\tilde{\omega}) = 
&- \frac{N_{0}}{2|\tilde{t}_{1}|}
\left[
4\tilde{g}_{0}(\tilde{g}_{1}-2\tilde{g}_{2})\,\epsilon_{k}
+ 8\tilde{g}_{0}\tilde{g}_{2}\,X_{+} - \frac{ P(k,X_{+})}{\sqrt{X_{+}^{2}-1}}\Theta\left(|X_{+}|-1\right)
\right] \notag \\
&- \frac{N_{0}+1}{2|\tilde{t}_{1}|}
\left[
4\tilde{g}_{0}(\tilde{g}_{1}-2\tilde{g}_{2})\,\epsilon_{k}
+ 8\tilde{g}_{0}\tilde{g}_{2}\,X_{-} - \frac{ P(k,X_{-})}{\sqrt{X_{-}^{2}-1}}\Theta\left(|X_{-}|-1\right)
\right]
\end{align}
where
\begin{align}
P(k,X) = \left( \tilde{g}_{0}^{2} - 4\tilde{g}_{0}\tilde{g}_{1} - 4\tilde{g}_{0}\tilde{g}_{2} + 4\tilde{g}_{0}\tilde{g}_{2}\epsilon_{2k} \right) \sgn(X-1) \notag
+ 4\tilde{g}_{0}(\tilde{g}_{1}-2\tilde{g}_{2})\,\epsilon_{k} \, \sgn(X+1) \,X
+ 8\tilde{g}_{0}\tilde{g}_{2} \, \sgn(X-1) \,X^{2} ~.
\end{align}
The analytical expression of the self-energy is depicted in Fig.\,\ref{Appendix SelfEnergy} and compared to the numerical evaluation.

\subsection{Non-adiabatic limit $|\tilde{t}_{1}| \ll \tilde{\omega}_{0}$}
In the non-adiabatic limit $|\tilde{t}_{1}| \ll \tilde{\omega}_{0}$, it is possible to analytically determine the binding energy and the effective mass of the polaron. In particular, we aim to discuss the vibrational modes change modify the dispersion relation of the electron. So we consider the case $\tilde{\omega},\, |\tilde{t}_{1}| \ll \tilde{\omega}_{0}$ (and $\tilde{g}_{0} \ll |\tilde{t}_{1}| $ for we consider the weak-coupling regime), which implies $|X_{\pm}| \gg 1$ and
\begin{align}
a \pm b |X_{\pm}| \mp \frac{c \pm a |X_{\pm}| + b |X_{\pm}|^{2}}{\sqrt{|X_{\pm}|^{2}-1}} \simeq \mp \frac{c+b/2}{|X_{\pm}|} ~.
\end{align}
This relation can be used to evaluate $\Real \tilde{\Sigma}^{(2)}_{R}$ in Eq.\,(\ref{Appendix Real Self Energy Single Electron}) and it leads to
\begin{align}
\Real \tilde{\Sigma}^{(2)}_{R}(k,\tilde{\omega}) = 
& - \tilde{\Delta}(k) - \left( 2N_{0} + 1 \right) \frac{\tilde{\Delta(k)}}{\tilde{\omega}_{0}} \tilde{\omega} ~,
\end{align}
where
\begin{align}
\tilde{\Delta}(k) = \frac{\tilde{g}_{0}^{2} - 4\tilde{g}_{0}\tilde{g}_{1}+4\tilde{g}_{0}\tilde{g}_{2}\epsilon_{2k}}{\tilde{\omega}_{0}} ~.
\end{align}
In order to determine the effective mass of the polaron, we assume that the single electron is associated to the following parabolic dispersion relation
\begin{align}
\epsilon_{k,0} \simeq \frac{k^{2}}{2\tilde{m}} ~,
\end{align}
where it is implied that the electron mass already takes into account the band flattening induced by the time-periodic driving. So it depends on the driving strength in the following way
\begin{align}
\tilde{m}(z) = \frac{1}{\tilde{t}_{1}(z)} = \frac{1}{\tilde{\nu} J_{0}(z)} ~.
\end{align}
In the limit of small $k$, the polaron dispersion relation $\xi_{k}$ is well described by
\begin{align}
\tilde{\xi}_{k} &= \epsilon_{k,0} + \Real \tilde{\Sigma}^{(2)}_{R}(k,\tilde{\xi}_{k}) \notag \\
&\simeq -\tilde{\Delta} + \frac{1}{1+ (2N_{0}+1) \frac{\tilde{\Delta}}{\tilde{\omega}_{0}}} \, \frac{k^{2}}{2\tilde{m}} ~,
\end{align}
where the polaron binding energy is
\begin{align}
\tilde{\Delta} = \frac{\tilde{g}_{0}^{2} - 4\tilde{g}_{0}\tilde{g}_{1}+4\tilde{g}_{0}\tilde{g}_{2}}{\tilde{\omega}_{0}} ~,
\end{align}
and the effective mass of the polaron satisfies
\begin{align}
\frac{\tilde{m}^{*}}{\tilde{m}} = 1+ (2N_{0}+1)\frac{\tilde{\Delta}}{\tilde{\omega}_{0}} ~.
\end{align}
When the off-resonant driving is turned off, i.e. when $z=0$, the binding energy reduces to $\tilde{\Delta} = \tilde{g}_{0}^{2}/\tilde{\omega}_{0}$ and
the expressions above provide the well-known results obtained in equilibrium.

\subsection{Adiabatic limit $\tilde{\omega}_{0} \ll |\tilde{t}_{1}|$}
In the adiabatic limit $\tilde{\omega}_{0} \ll |\tilde{t}_{1}|$, it is also possible to analytically characterize the binding energy and the effective mass of the polaron. We consider two cases:

\begin{itemize}
\item $-2\tilde{t}_{1} - \tilde{\omega}_{0} < \tilde{\omega} < - 2\tilde{t}_{1} + \tilde{\omega}_{0}$, which corresponds to $|X_{+}|<1$ and $|X_{-}|>1$. This allows us to study polaron properties for energies in the vicinity of $-2\tilde{t}_{1}$, which corresponds to the bottom (top) of the band when $\tilde{t}_{1}(z)>0~(<0)$. To do so, we assume $\tilde{\omega} = -2\tilde{t}_{1} + h$ with $\tilde{t}_{1}>0$ and $h \ll \tilde{\omega}_{0} \ll \tilde{t}_{1}$ such that
\begin{align}
\Real \tilde{\Sigma}^{(2)}_{R}(k,\tilde{\omega})
\simeq &-\frac{N_{0}}{2|\tilde{t}_{1}|} \left[ a - b + b \frac{h+\tilde{\omega}_{0}}{2\tilde{t}_{1}} \right] \notag \\
&-\frac{N_{0}+1}{2|\tilde{t}_{1}|} \left[ a - b + b \frac{h-\tilde{\omega}_{0}}{2\tilde{t}_{1}} + \sqrt{\frac{2|\tilde{t}_{1}|}{2(\tilde{\omega}_{0}-h)}} \left( c - a - b - (a-2b) \frac{\tilde{\omega}_{0}-h}{2\tilde{t}_{1}} \right) \right] \\
\simeq &-\left( N_{0} + 1 \right) \frac{c-a-b}{\sqrt{4|\tilde{t}_{1}|\tilde{\omega_{0}}}}\left( 1 + \frac{h}{2\tilde{\omega}_{0}} \right)
\end{align}
The explicit expressions of coefficients $a$, $b$ and $c$ can be found from Eq.\,(\ref{Appendix Real Self Energy Single Electron}). Again we assume that the single electron is characterized by the parabolic dispersion relation
\begin{align}
\epsilon_{k,0} \simeq \frac{k^{2}}{2\tilde{m}} ~.
\end{align}
The definition of the driving-renormalized electron mass has already been introduced above. This results in
\begin{align}
\tilde{\xi}_{k} &\simeq -\tilde{\Delta} + \frac{1}{1+ \frac{\tilde{\Delta}}{2\tilde{\omega}_{0}}} \frac{k^{2}}{2\tilde{m}} ~,
\end{align}
where the polaron binding energy is
\begin{align}
\tilde{\Delta} = (N_{0}+1) \, \frac{\tilde{g}_{0}^{2} - 8\tilde{g}_{0}\tilde{g}_{1}}{\sqrt{4\tilde{t}_{1}\tilde{\omega}_{0}}} ~,
\end{align}
and the effective mass of the polaron satisfies
\begin{align}
\frac{\tilde{m}^{*}}{\tilde{m}} = 1+ \frac{\tilde{\Delta}}{2\tilde{\omega}_{0}} ~.
\end{align}
When the off-resonant driving is turned off the binding energy also reduces to $\tilde{\Delta} = \tilde{g}_{0}^{2}/\tilde{\omega}_{0}$ and
the expressions above provide the well-known results obtained in equilibrium.

\item $-\tilde{t}_{1} + \tilde{\omega}_{0} < \tilde{\omega} < \tilde{t}_{1} - \tilde{\omega}_{0}$, which corresponds to $|X_{+}|<1$ and $|X_{-}|<1$. This describes almost all energies within the electron band (remember that $\tilde{\omega}_{0} \ll \tilde{t}_{1}$), except the vicinities of top and bottom which are described in the previous case. Then Eq.\,(\ref{Appendix Real Self Energy Single Electron}) directly leads to
\begin{align}
\Real \tilde{\Sigma}^{(2)}_{R}(k,\tilde{\omega})
&= 
- \frac{2N_{0}+1}{|\tilde{t}_{1}|^{2}}
\left[
\tilde{g}_{0}(\tilde{g}_{1}-2\tilde{g}_{2})\,\epsilon_{k,0}
+ 2\tilde{g}_{0}\tilde{g}_{2}\, \tilde{\omega}
\right]
+ 2\frac{\tilde{g}_{0}\tilde{g}_{2}}{\tilde{t}_{1}^{2}} \, \tilde{\omega}_{0} ~.
\end{align}
The polaronic band is given by
\begin{align}
\tilde{\xi}_{k} &=  \tilde{\Delta} + \frac{1}{2\tilde{m}^{*}} \, \epsilon_{k} ~.
\end{align}
where the binding energy is
\begin{align}
\tilde{\Delta} = 2\frac{\tilde{g}_{0}\tilde{g}_{2}}{\tilde{t}_{1}^{2}} \, \tilde{\omega}_{0}
\end{align}
and the effective mass satisfies
\begin{align}
\frac{\tilde{m}^{*}}{\tilde{m}} = 1+ (2N_{0}+1) \frac{\tilde{g}_{0}\tilde{g}_{1}}{\tilde{t}_{1}^{2}} ~.
\end{align}
Note that we have not made any assumptions upon $\tilde{\omega}$. Thus the expressions above describe all energies smaller than $|\tilde{t}_{1}-\tilde{\omega}_{0}|$. In other words, it has been possible to obtain the exat expression of the polaron band for all values of $k$ in the non-adiabatic limit. Moreover, the electron has more energy than the phonon frequency, so it is also allowed to emit a phonon, even at zero temperature when $N_{0}=0$. This yields a nonzero imaginary part to the self-energy. The zeroth order in the limit $\tilde{\omega}_{0} \ll \tilde{y}_{1}$ leads to a polaron life time $\tau$ that satisfies
\begin{align}
\frac{1}{\tau(k,\tilde{\omega})} &=
- \Imag \tilde{\Sigma}^{(2)}_{R}(k,\tilde{\omega}) \notag \\
&=
\left[
\tilde{g}_{0}^{2} - 4\tilde{g}_{0}\tilde{g}_{1} - 4\tilde{g}_{0}\tilde{g}_{2} + 4\tilde{g}_{0}\tilde{g}_{2}\epsilon_{2k} 
+ 4\tilde{g}_{0}(\tilde{g}_{1}-2\tilde{g}_{2})\,\epsilon_{k}\,X 
+ 8\tilde{g}_{0}\tilde{g}_{2}\,X^{2}
\right] \, 
\frac{2N_{0}+1}{\sqrt{1-X^{2}}} ~.
\end{align}
The quasiparticle lifetime is already finite in equilibrium. However nonequilibrium corrections make it $k$-dependent. 
\end{itemize}

%%%
\section{Lang-Firsov canonical transformation}\label{Appendix Lang-Firsov canonical transformation}
We start from the following effective Hamiltonian:
\begin{align}
\tilde{H} = \tilde{t}_{1} \sum_{m} \left( c^{\dagger}_{m+1}c_{m}+h.c. \right) + \sum_{q} \tilde{\omega}_{q} \, b^{\dagger}_{q}b_{q} 
+ \sum_{m,q} \tilde{g}_{q} \, e^{-iqm} \, c^{\dagger}_{m}c_{m}B_{q}
+ \tilde{g}_{2} \sum_{m,q} \beta_{q} e^{-iqm} \left( c^{\dagger}_{m+2}c_{m}+h.c. \right) B_{q} ~,
\end{align}
where $\tilde{g}_{q} = \tilde{g}_{0} + \alpha_{q} \tilde{g}_{1}$, $\alpha_{q} = 2(\epsilon_{q} - 1)$ and $\beta_{q} = 1 - 2e^{-iq} + e^{-i2q}$. The standard Lang-Firsov transformation consists of
\begin{align}
H=e^{S}\tilde{H}e^{-S} ~,
\end{align}
with
\begin{align}
S = -\sum_{mq} u_{q} \, e^{-iqm}c_{m}^{\dagger}c_{m} \, (b_{q}-b_{-q}^{\dagger}) ~~~~~~ \text{and} ~~~~~~ u_{q} = \frac{\tilde{g}_{q}}{\tilde{\omega}_{q}} ~.
\end{align}
It transforms bosonic and fermionic operators according to
\begin{align}
e^{S}b_{q}e^{-S} = b_{q} - \sum_{m}u_{q} \, e^{iqm}c_{m}^{\dagger}c_{m} ~~~~~~ \text{and} ~~~~~~ e^{S}c_{m}e^{-S} = c_{m}X_{m} ~,
\end{align}
where the operator $X_{m}$ is defined as
\begin{align}
X_{m} = \exp \left( \sum_{q} u_{q} \, e^{-iqm} (b_{q}-b_{-q}^{\dagger}) \right)
\end{align}
and commutes with fermionic operators. The transformation turns the effective Hamiltonian into
\begin{align}
\tilde{H}' &= \sum_{q} \tilde{\omega}_{q} b_{q}^{\dagger}b_{q} - \sum_{mnq}\frac{\tilde{g}_{q}^{2}}{\tilde{\omega}_{q}}e^{-iq(m-n)}c^{\dagger}_{m}c_{m}c^{\dagger}_{n}c_{n} \notag \\
&+ \tilde{g}_{2} \sum_{mq} \beta_{q} e^{-iqm} c^{\dagger}_{m+2}c_{m}X_{m+2}^{\dagger}X_{m}\left( b_{q}+b_{-q}^{\dagger} +2\sum_{n} u_{q} \, e^{iqn} c_{n}^{\dagger}c_{n} \right) \notag \\
&+ \tilde{g}_{2} \sum_{mq} \beta_{q} e^{-iqm} c^{\dagger}_{m}c_{m+2}X_{m}^{\dagger}X_{m+2}\left( b_{q}+b_{-q}^{\dagger} +2\sum_{n} u_{q} \, e^{iqn} c_{n}^{\dagger}c_{n} \right) \notag \\
%&= \sum_{q}\tilde{\omega}_{q} b_{q}^{\dagger}b_{q} - \sum_{mnq}\frac{\left[\tilde{g}_{0}+\tilde{g}_{1}\alpha_{q}\right]^{2}}{\tilde{\omega}_{q}}e^{-iq(m-n)}c^{\dagger}_{m}c_{m}c^{\dagger}_{n}c_{n} \notag \\
%&+ \tilde{g}_{2} \sum_{mq} \beta_{q} e^{-iqm} c^{\dagger}_{m+2}c_{m}X_{m+2}^{\dagger}X_{m}\left( b_{q}+b_{-q}^{\dagger} \right)
%+ 2\tilde{g}_{2} \sum_{mnq} \beta_{q} \, u_{q} \, e^{-iq(m-n)} c^{\dagger}_{m+2}c_{m}c_{n}^{\dagger}c_{n}X_{m+2}^{\dagger}X_{m} \notag \\
%&+ \tilde{g}_{2} \sum_{mq} \beta_{q} e^{-iqm} c^{\dagger}_{m}c_{m+2}X_{m}^{\dagger}X_{m+2}\left( b_{q}+b_{-q}^{\dagger} \right)
%+ 2\tilde{g}_{2} \sum_{mnq} \beta_{q} \, u_{q} \, e^{-iq(m-n)} c^{\dagger}_{m}c_{m+2}c_{n}^{\dagger}c_{n}X_{m}^{\dagger}X_{m+2} \notag \\
&= \sum_{q}\tilde{\omega}_{q} b_{q}^{\dagger}b_{q} - \tilde{\Delta} \sum_{q}c^{\dagger}_{m}c_{m}
+ \tilde{t}_{1} \sum_{m} \left( c^{\dagger}_{m+1}c_{m}X_{m+1}^{\dagger}X_{m} + h.c. \right)
+ \tilde{t}_{2} \sum_{m} \left( c^{\dagger}_{m+2}c_{m}X_{m+2}^{\dagger}X_{m} + h.c. \right) \notag \\
&+ \tilde{g}_{2} \sum_{mq} \beta_{q} e^{-iqm} \left( c^{\dagger}_{m+2}c_{m}X_{m+2}^{\dagger}X_{m} + h.c. \right) B_{q} ~,
\end{align}
where interactions between polarons have been neglected, $\tilde{\Delta} = \sum_{q} \tilde{g}_{q}^{2}/\tilde{\omega}_{q}$ and $\tilde{t}_{2} = 2\tilde{g}_{2} \sum_{q} \beta_{q} u_{q}$. For dispersionless phonons of frequency $\tilde{\omega}_{0}$, the onsite energy is given by
\begin{align}
\tilde{\Delta} &= \frac{1}{\tilde{\omega}_{0}} \sum_{q} \left[\tilde{g}_{0}+\tilde{g}_{1}\alpha_{q}\right]^{2} \notag \\
&\simeq \frac{1}{\tilde{\omega}_{0}} \sum_{q} \left(\tilde{g}_{0}^{2}+2\tilde{g}_{0}\tilde{g}_{1}\alpha_{q}\right) \notag \\
&\simeq \frac{1}{\tilde{\omega}_{0}} \sum_{q} \left(\tilde{g}_{0}^{2}+4\tilde{g}_{0}\tilde{g}_{1}(\cos q -1)\right) \notag \\
&\simeq \frac{\tilde{g}_{0}^{2}-4\tilde{g}_{0}\tilde{g}_{1}}{\tilde{\omega}_{0}} ~,
\end{align}
and
\begin{align}
\tilde{t}_{2} &= 2\tilde{g}_{2}\sum_{q}\beta_{q}u_{q} \notag \\
&= 2\tilde{g}_{2}\sum_{q} (1 - 2e^{-iq} + e^{-i2q})\frac{\tilde{g}_{0}+2\tilde{g}_{1}(\cos q -1)}{\tilde{\omega}_{0}} \notag \\
%&= 2\tilde{g}_{2}\frac{\tilde{g}_{0} - 4\tilde{g}_{1}}{\tilde{\omega}_{0}} ~.
&\simeq 2\frac{\tilde{g}_{0}\tilde{g}_{2}}{\tilde{\omega}_{0}} ~.
\end{align}
In the expressions above we have neglected products like $\tilde{g}_{1}\tilde{g}_{2}$, since they yield a $\lambda^{4}$ contribution.

%%%
\section{Peierls-Feynman-Bogolyubov variational principal}\label{Appendix Peierls-Feynman-Bogolyubov variational principal}
The canonical transformation does not diagonalize the effective Hamiltonian and yields a nonlocal interaction between a next-nearest-neighbor hopping polaron and the lattice vibrations it feels along it motion, which makes the analytical description complicated \textit{a priori}. In order to overcome this complexity, we aim to map Hamiltonian $\tilde{H}'$ onto
\begin{align}
H^{*} &= \sum_{q}\tilde{\omega}_{q} \, b_{q}^{\dagger}b_{q} 
- \tilde{\Delta} \sum_{m}c^{\dagger}_{m}c_{m}
+ t_{1}^{*} \sum_{m} \left(c^{\dagger}_{m+1}c_{m}+h.c.\right)  + t_{2}^{*} \sum_{m} \left(c^{\dagger}_{m+2}c_{m}+h.c.\right)
\end{align}
This Hamiltonian is quadratic in momentum space, so that we know its partition function $Z^{*}=\Tr e^{-\beta H^{*}}$. Parameters $t_{1}^{*}$ and $t_{2}^{*}$ are then determined under the constraint that $\rho^{*} = \Tr e^{-\beta H^{*}}/Z^{*}$ is the best approximation of the exact density operator defined from Hamiltonian $\tilde{H}'$. This leads to Peierls-Feynman-Bogoliubov variational principle which consists in minimizing with respect to $t_{1}^{*}$ and $t_{2}^{*}$ the functional
\begin{align}
F^{*}+\langle \tilde{H}' - H^{*} \rangle_{*} ~,
\end{align}
where $F^{*} = - (1/\beta) \ln Z^{*}$. This requires the calculation of the following average
\begin{align}
\langle \tilde{H}' - H^{*} \rangle_{*} 
&= 
\tilde{t}_{1} \sum_{m} \left\langle c^{\dagger}_{m+1}c_{m} \right\rangle_{*} \left\langle X_{m+1}^{\dagger}X_{m} \right\rangle_{*}
- t_{1}^{*} \sum_{m} \left\langle c^{\dagger}_{m+1}c_{m} \right\rangle_{*} \notag \\
&+ \tilde{t}_{2} \sum_{m} \left\langle c^{\dagger}_{m+2}c_{m} \right\rangle_{*} \left\langle X_{m+2}^{\dagger}X_{m} \right\rangle_{*}
- t_{2}^{*} \sum_{m} \left\langle c^{\dagger}_{m+2}c_{m} \right\rangle_{*} \notag \\
&+ \tilde{g}_{2} \sum_{mq} \beta_{q} e^{-iqm} \left\langle c^{\dagger}_{m+2}c_{m}\right\rangle_{*} \left\langle X_{m+2}^{\dagger}X_{m}\,B_{q} \right\rangle_{*} \notag \\
&+h.c.
\end{align}
where
\begin{align}
X_{m}^{\dagger}X_{n} = \exp \left( \sum_{q} u_{q} \, (e^{-iqn}-e^{-iqm})(b_{q}-b_{-q}^{\dagger}) \right) ~.
\end{align}
The average of this bosonic operator can be estimated via Feynman disentangling method as follows
\begin{align}
\left\langle X_{m+n}^{\dagger}X_{m} \right\rangle_{*} 
&= \left\langle e^{\sum_{q}(v_{m,q}b_{q}-v^{*}_{m,q}b_{q}^{\dagger})} \right\rangle_{*} \notag \\
&= \prod_{q \in BZ} \left(1-e^{-\beta\tilde{\omega}_{q}}\right) \sum_{n_{q}=0}^{+\infty} e^{-\beta \tilde{\omega}_{q} n_{q}} \langle n_{q } | e^{(v_{m,q}b_{q}-v^{*}_{m,q}b_{q}^{\dagger})} | n_{q} \rangle \notag \\
&= \prod_{q \in BZ} e^{-\frac{v_{m,q}v_{m,q}^{*}[b_{q},b_{q}^{\dagger}]}{2}} \left(1-e^{-\beta\tilde{\omega}_{q}}\right) \sum_{n_{q}=0}^{+\infty} e^{-\beta \tilde{\omega}_{q} n_{q}} \langle n_{q } | e^{v_{m,q}b_{q}}e^{-v^{*}_{m,q}b_{q}^{\dagger}} | n_{q} \rangle \notag \\
&= \prod_{q \in BZ} e^{-\frac{|v_{m,q}|^{2}}{2}} \left(1-e^{-\beta\tilde{\omega}_{q}}\right) \sum_{n_{q}=0}^{+\infty} e^{-\beta \tilde{\omega}_{q} n_{q}} \sum_{m=0}^{+\infty} \frac{(-|v_{m,q}|^{2})^{m}}{(m!)^{2}} \frac{n_{q}!}{(n_{q}-m)!} \notag \\
&= \prod_{q \in BZ} e^{-\frac{|v_{m,q}|^{2}}{2}} \left(1-e^{-\beta\tilde{\omega}_{q}}\right) \sum_{n_{q}=0}^{+\infty} e^{-\beta \tilde{\omega}_{q} n_{q}} \, L_{n_{q}}(|v_{m,q}|^{2}) \notag \\
&= \prod_{q \in BZ} e^{-\frac{|v_{m,q}|^{2}}{2}} e^{-|v_{m,q}|^{2}N_{q}} \notag \\
&= \prod_{q \in BZ} e^{-|v_{m,q}|^{2}(N_{q}+\frac{1}{2})}
\end{align}
where $L_{n_{q}}$ denotes the Laguerre polynomial of order $n$, $N_{q}$ is the equilibrium distribution function that characterizes phonons of frequency $\tilde{\omega}_{q}$. Besides
\begin{align}
v_{m,q} = u_{q} \, e^{-iqm}(1-e^{-inq}) ~,
\end{align}
so that
\begin{align}
|v_{m,q}|^{2} = 2 u_{q}^{2} (1-\cos nq) ~,
\end{align}
and finally
\begin{align}
\left\langle X_{m+n}^{\dagger}X_{m} \right\rangle_{*} &= \prod_{q \in BZ} e^{-u_{q}^{2}(1-\cos nq)(2N_{q}+1)} = \left\langle X_{m}^{\dagger}X_{m+n} \right\rangle_{*}~.
\end{align}
It is worth mentioning that this average does not depend on atomic coordinate $m$, but it does depend on interatomic distance $n$.
Another average which still has to be evaluated is
\begin{align}
\left\langle X_{m+2}^{\dagger}X_{m}\left( w_{q',m} \, b_{q'} + w_{m,q'}^{*} b_{-q'}^{\dagger} \right) \right\rangle_{*}
&= \left. \partial_{\phi} \left\langle X_{m+2}^{\dagger}X_{m} e^{\phi\left( w_{q',m} \, b_{q'} + w_{m,q'}^{*} b_{-q'}^{\dagger} \right)} \right\rangle_{*} \right|_{\phi=0} \notag \\
&= \partial_{\phi} \prod_{q \in BZ} e^{\frac{(v_{m,q}w_{m,q}^{*}-v_{m,q}^{*}w_{m,q})\delta_{qq'}}{2}\phi} \left(1-e^{-\beta\omega_{q}}\right) \\
&\times \left. \sum_{n_{q}=0}^{+\infty} e^{-\beta \omega_{q} n_{q}} \langle n_{q } | e^{\left( (v_{m,q}+\phi w_{m,q}\delta_{q,q'})b_{q}-(v^{*}_{m,q}- \phi w_{m,q}^{*}\delta_{qq'})b_{q}^{\dagger}\right)} \left( b_{q'}+b_{-q'}^{\dagger} \right) | n_{q} \rangle \right|_{\phi=0} \notag \\
&= \left. \partial_{\phi} 
\prod_{q \in BZ} e^{\frac{(v_{m,q}w_{m,q}^{*}-v_{m,q}^{*}w_{m,q})\delta_{qq'}}{2}\phi}
e^{-(v_{m,q}+\phi w_{m,q}\delta_{q,q'})(v^{*}_{m,q}-\phi w_{m,q}^{*}\delta_{qq'})(N_{q}+\frac{1}{2})}
\right|_{\phi=0} \notag \\
&= 2i\Imag[v_{m,q'}w_{m,q'}^{*}] (N_{q'}+1) \left\langle X_{m+2}^{\dagger}X_{m} \right\rangle_{*}
\end{align}
where $w_{m,q}=\beta_{q} \, e^{-iqm}$. As far as we are concerned, phonons are dispersionless so that $\omega_{q}=\omega_{0}$ and $N_{q}=N_{0}$. As a result
\begin{align}
\left\langle X_{m+n}^{\dagger}X_{m} \right\rangle_{*}
&= \exp \left( - (2N_{0}+1) \sum_{q} u_{q}^{2} \, (1-\cos nq) \right) \notag \\
&= \exp \left( - (2N_{0}+1) \sum_{q} \left( \frac{\tilde{g}_{0}+\tilde{g}_{1}\alpha_{q}}{\tilde{\omega}_{0}} \right)^{2} \, (1-\cos nq) \right) \notag \\
&= \exp \left( - (2N_{0}+1) \sum_{q} \left( \frac{\tilde{g}_{0}+2\tilde{g}_{1}(\cos q - 1 )}{\tilde{\omega}_{0}} \right)^{2} \, (1-\cos nq) \right) \notag \\
&\simeq \exp \left( - \frac{2N_{0}+1}{\tilde{\omega}_{0}^{2}} \sum_{q} \left( \tilde{g}_{0}^{2}+4\tilde{g}_{0}\tilde{g}_{1}(\cos q - 1 ) \right) \, (1-\cos nq) \right) \notag \\
&\simeq \exp \left( - (2N_{0}+1)\frac{\tilde{g}_{0}^{2}-4\tilde{g}_{0}\tilde{g}_{1}-2\tilde{g}_{0}\tilde{g}_{1}\delta_{n,1}}{\tilde{\omega}_{0}^{2}} \right)
\end{align}
and
\begin{align}
\sum_{q} \Imag[v_{m,q}w_{m,q}^{*}]
&= \sum_{q} \Imag[u_{q} \, e^{-iqm}(1-e^{-i2q}) \, \beta_{q}^{*} \, e^{+iqm}] \notag \\
&= \sum_{q} \Imag \left[ \frac{\tilde{g}_{0}+\tilde{g}_{1}\alpha_{q}}{\tilde{\omega}_{0}}(1-e^{-i2q}) (1 - 2e^{+iq} + e^{+i2q}) \right] \notag \\
&= \sum_{q} \Imag \left[ 4i \frac{\tilde{g}_{0}+2\tilde{g}_{1}(\cos q -1)}{\tilde{\omega}_{0}}\sin2q \, (\cos q -1) \right] \notag \\
&= 4\sum_{q} \frac{\tilde{g}_{0}+2\tilde{g}_{1}(\cos q -1)}{\tilde{\omega}_{0}}\sin2q \, (\cos q -1) \notag \\
&= 0 
\end{align}
since this relies on the integral of an odd function of $q$. After introducing
\begin{align}
t_{1} &= \tilde{t}_{1} \left\langle X_{m+1}^{\dagger}X_{m} \right\rangle_{*} = \tilde{t}_{1} \exp \left( - (2N_{0}+1)\frac{\tilde{g}_{0}^{2}-6\tilde{g}_{0}\tilde{g}_{1}}{\tilde{\omega}_{0}^{2}} \right)
\end{align}
and
\begin{align}
t_{2} &= \tilde{t}_{2} \left\langle X_{m+2}^{\dagger}X_{m} \right\rangle_{*} = \tilde{t}_{2} \exp \left( - (2N_{0}+1)\frac{\tilde{g}_{0}^{2}-4\tilde{g}_{0}\tilde{g}_{1}}{\tilde{\omega}_{0}} \right) ~,
\end{align}
we end up with the following expression
\begin{align}
\langle \tilde{H}' - H^{*} \rangle_{*} 
&= 
\left( t_{1} - t_{1}^{*} \right) \sum_{m} \left\langle c^{\dagger}_{m+1}c_{m} + h.c. \right\rangle_{*} + \left( t_{2} - t_{2}^{*} \right) \sum_{m} \left\langle c^{\dagger}_{m+2}c_{m} + h.c. \right\rangle_{*} ~.
\end{align}
Minimizing the functional $F^{*}+\langle \tilde{H}' - H^{*} \rangle_{*}$ with respect to $t_{1}^{*}$ and $t_{2}^{*}$ then leads to
\begin{align}\label{Appendix Minimizing Condition}
\left\{
\begin{aligned}
\sum_{k} \left[\left( t_{1} - t_{1}^{*} \right)\,\epsilon_{k} + \left( t_{2} - t_{2}^{*} \right)\,\epsilon_{2k}\right] \, \partial_{t_{1}^{*}} \, \left\langle c^{\dagger}_{k}c_{k} \right\rangle_{*} = 0 \\
\sum_{k} \left[\left( t_{1} - t_{1}^{*} \right)\,\epsilon_{k} + \left( t_{2} - t_{2}^{*} \right)\,\epsilon_{2k}\right] \, \partial_{t_{2}^{*}} \, \left\langle c^{\dagger}_{k}c_{k} \right\rangle_{*} = 0
\end{aligned} \right.
\end{align}
where
\begin{align}
\left\langle c^{\dagger}_{k}c_{k} \right\rangle_{*} &= 1/(1+e^{\beta( t_{1}^{*}\epsilon_{k}+t_{2}^{*}\epsilon_{2k})}) \notag \\
&=n_{k} ~.
\end{align}
The system of Eq\,(\ref{Appendix Minimizing Condition}) implies
\begin{align}
\sum_{k} \left[\left( t_{1} - t_{1}^{*} \right)\,\epsilon_{k} + \left( t_{2} - t_{2}^{*} \right)\,\epsilon_{2k}\right]^{2} \, n_{k}' = 0 ~.
\end{align}
Because $n' = \partial_{X} \left[ 1/(1+e^{\beta X})\right] < 0$, the functional $F^{*}+\langle \tilde{H}' - H^{*} \rangle_{*}$ is finally minimized when 
\begin{align}\label{Appendix Minimizing Condition}
\left\{
\begin{aligned}
t_{1}^{*} = t_{1} \\
t_{2}^{*} = t_{2}
\end{aligned} \right. ~.
\end{align}

\end{document}